\documentclass[manuscript]{acmart}

\AtBeginDocument{%
  \providecommand\BibTeX{{%
    \normalfont B\kern-0.5em{\scshape i\kern-0.25em b}\kern-0.8em\TeX}}}

\setcopyright{acmcopyright}
\copyrightyear{2023}
\acmYear{2023}
\acmDOI{XXXXXXX.XXXXXXX}

\acmConference[CSCW'26]{The ACM Conference on Computer-Supported Cooperative Work and Social Computing}{June 03--05, 2018}{Woodstock, NY}
\acmPrice{15.00}
\acmISBN{978-1-4503-XXXX-X/18/06}

\usepackage{graphicx}
\usepackage{svg}

\usepackage{subcaption}
\usepackage{pdfpages}
\usepackage{lineno}
\usepackage{enumitem}
\usepackage{booktabs}
\usepackage{caption}
\usepackage{array}

\usepackage[hide]{comments}
\Commenter{NT}

\begin{document}
\nolinenumbers

\title[From Inquisitorial to Adversarial]{From Inquisitorial to Adversarial: Using Legal Theory to Redesign Reporting Systems \revise{in Online Communities}}

\author{Leijie Wang}
\email{leijiew@cs.washington.edu}
\affiliation{%
  \institution{University of Washington}
  \city{Seattle}
  \country{United States}
}

\author{Weizi Wu}
\email{weiziw@uw.edu}
\affiliation{%
  \institution{University of Washington}
  \city{Seattle}
  \country{United States}
}

\author{Lirong Que}
\email{lq22@cs.washington.edu}
\affiliation{%
  \institution{University of Washington}
  \city{Seattle}
  \country{United States}
}

\author{Nirvan Tyagi}
\email{tyagi@cs.washington.edu}
\affiliation{%
  \institution{University of Washington}
  \city{Seattle}
  \country{United States}
}

\author{Amy X. Zhang}
\email{axz@cs.uw.edu}
\affiliation{%
  \institution{University of Washington}
  \city{Seattle}
  \country{United States}
}

\newcommand\revise[1]{#1}
\newcommand\revisecomment[2]{#2}
\newcommand\move[2]{#2}

\renewcommand{\shortauthors}{Wang Leijie et al.}


\begin{abstract}
\revise{In online communities, user reporting systems play a central role in addressing interpersonal conflict and online harassment, particularly in spaces with heightened expectations of privacy, such as direct messages, voice chats, and end-to-end encrypted messaging.
These private spaces complicate how community moderators collect evidence and intensify user concerns about procedural justice and privacy.
To examine these challenges, we apply adversarial legal frameworks from offline judicial systems to community-level reporting systems, using Discord as a research site.
We find that reporting systems in online communities tend to follow an \textit{inquisitorial model}, where moderators lead evidence collection and case development, in contrast to \textit{adversarial models} that grant users greater control.
While adversarial practices can enhance procedural justice and protect privacy, they also introduce new risks of abuse, underscoring the need for careful threat modeling.
Building on this analysis, we present a design space that grants users greater control over disclosing and authenticating evidence while accounting for the privacy constraints and technical affordances of online communities.
We conclude by discussing how this design space could inform platform-level reporting systems and be reinforced through cryptographic techniques amid growing distrust in platforms.}
\end{abstract}

\ccsdesc[500]{Human-centered computing~Collaborative and social computing systems and tools}

\keywords{}

\maketitle
\margincommentprimer

\section{Introduction}
\revise{Online communities such as Discord servers, Slack workspaces, Facebook groups, and WhatsApp groups have become central sites of everyday social interaction, ranging from casual conversations among internet strangers to more personal exchanges among collaborators and close friends~\cite{pew2025socialmedia, discord2025resourcera}.
Many of these interactions now take place in \textit{private spaces}, where participants have a heightened expectation of privacy. 
Examples of private spaces include direct messages, private channels~\cite{silabPrivateChatAnalytics}, voice chats~\cite{jiang2019moderation}, ephemeral messages~\cite{kotfila2014message}, and end-to-end encrypted (E2EE) messaging~\cite{kamara2022outside, pfefferkorn2022content}.}
However, interactions in these spaces can sometimes escalate into interpersonal conflict~\cite{levy2022understanding, im2018deliberation} or online harassment~\cite{thomas2021sok}.
While community moderators often rely on automated systems to detect problematic behavior~\cite{jhaver2019human, Chandrasekharan2019}, such tools are typically inappropriate in these private spaces because they represent automated monitoring of private conversations.
As a result, user reporting systems remain the primary means of surfacing problematic behavior in online communities~\cite{pfefferkorn2022content, kamara2022outside}.

In a typical community-level reporting system, users can report others by flagging and sharing content they consider inappropriate with community moderators.
Moderators then review the report, gather evidence, and decide whether to punish the accused~\cite{jhaver2019human, seering2019moderator}.
\revise{Evidence collection plays a central role in shaping how reports are interpreted and resolved.
Common forms of evidence include account information (e.g., join date, history of reporting or being reported) and user identifiers (to recall their reputations within communities)~\cite{cai2021after}.
Most critically, moderators often examine conversation logs that directly document the reported interactions~\cite{cai2021after, jiang2019moderation}, which we refer to as \textit{documentary evidence}.
To collect and verify this evidence, moderators may also reach out to the reporting user, the reported user, or bystanders~\cite{jiang2019moderation}.
Since most community moderators are part-time or volunteers~\cite{seering2019moderator}, community-level reporting systems tend to be informal and loosely structured.
Consequently, moderators often have broad discretion over how evidence is collected and how reports are resolved, mirroring the hierarchical power structures of many online communities~\cite{zhang2020policykit, schneider2022admins}.}

Such inherent power imbalance between moderators and reporting users contributes to recurring frustrations when reporting incidents that occur in private spaces~\cite{wang2023reporting, jiang2019moderation}.
In these situations, users often face a dilemma: either share more of their private conversation to support their report or risk bad outcomes due to a lack of compelling evidence~\cite{wang2023reporting}.
This tension is particularly challenging when proving harm requires revealing sensitive content, such as non-consensual intimate imagery~\cite{sultana2021unmochon, batool2024expanding}.
Such practices can further undermine users’ perceptions of procedural justice~\cite{kou2021flag, shouting2023pen}. Reporting systems often feel like black boxes, where users submit reports and later receive decisions with little insight into what evidence was collected or how it was evaluated.
Beyond transparency, users typically lack control over how evidence is gathered, including which moderators are involved or whom they may contact during the investigation.

To address these failures of online reporting systems, we draw inspiration from offline judicial systems, which have long explored procedural designs for balancing fairness, privacy, and abuse prevention~\cite{braithwaite2002restorative, daicoff2006law, freiberg2001problem}.
\revise{\textbf{Our first contribution is to apply adversarial legal framework~\cite{damaska1972evidentiary, damaska1986faces} as a lens to examine how community-level reporting systems collect evidence.}
We focused on Discord in our analysis because it hosts numerous communities~\cite{discord2025resourcera} and supports diverse private communication features~\cite{discord2025blockingprivacy, jiang2019moderation, discord2025messageapi, discord2025e2eeav}.
Our analysis draws from documentation in prior work and formative interviews with 5 moderators and 7 users on Discord.}
We find that most reporting systems follow \textit{an inquisitorial model}, in which judges play a central role in gathering evidence and shaping cases~\cite{damaska1986faces}.
By contrast, \textit{adversarial models} grant the involved parties greater control over how cases are investigated and what evidence is presented~\cite{damaska1972evidentiary, damaska1986faces}.
\revise{Applying this framework, we show how adversarial elements \footnote{Our use of “adversarial” draws from the adversarial models in comparative legal literature, and should not be confused with DiSalvo’s ``Adversarial Design'', which emphasizes agonism and political contestation in design research~\cite{disalvo2015adversarial}.} can enhance procedural justice and privacy but also introduce risks when parties exploit their control for self-interest~\cite{damaska1972evidentiary, kim2014adversarial}.
Therefore, we argue that incorporating adversarial elements into online reporting systems requires careful threat modeling that accounts for the unique constraints of online communities.
For instance, while offline courts can impose severe penalties to deter abuse~\cite{retaliating2025cornell}, this is less effective in online settings as users can often evade sanctions by creating new accounts or moving across platforms.
Consequently, we focused on designing techniques and interactions that embody adversarial practices while explicitly guarding against abuse.}

\revise{\textbf{This analysis motivates our second contribution: a rich design space for empowering users to collect and present documentary evidence while mitigating potential abuse of reporting procedures.}
We focus on documentary evidence specifically because it remains central to moderators’ decision-making, yet is often compromised by privacy-preserving features of online spaces~\cite{jiang2019moderation, wang2023reporting}.
To ground this design space, we introduce a threat model in which both users and moderators may misuse the reporting system but are constrained by what the user interface permits.}
Two core challenges emerge when granting users more control over evidence collection.
First, in \textbf{evidence disclosure}, users decide which conversations to share with moderators, raising a trade-off between protecting user privacy and providing moderators with enough information to make informed decisions.
Drawing on data minimization practices ~\cite{kulshrestha2021identifying, biega2020operationalizing, orekondy2018connecting}, we outline a design space that ensures shared conversations are adequate for, but limited to, moderation purposes.
Second, even if users do not selectively disclose their conversations, they still need to prove to moderators that their submitted private conversations happened, a process known in offline settings as \textbf{evidence authentication}~\cite{grimm2012authentication}.
While users often rely on screenshots for authentication purposes~\cite{sultana2021unmochon}, these methods are prone to abuse---for example, malicious users can forge conversations by mimicking usernames and avatars of another user.
We explore how an enhanced message forwarding functionality could enable platform-side authentication, and how web automation techniques can help users collect trustworthy evidence by preserving the chain of custody~\cite{pu2023dilogics, yeh2009sikuli}. 
However, because these approaches rely on persisted messages, they fall short in ephemeral contexts.
We propose \textit{ephemeral reporting} to extend accountability without compromising intended privacy.
\revise{We conclude by discussing how these ideas could inform platform-level reporting and be reinforced through cryptographic approaches in response to growing distrust in platforms~\cite{internetsociety2019privacy}.}
\section{Related Work}
\subsection{Moderation Mechanisms to Address Online Harassment and Interpersonal Conflicts}

\revise{Considerable research has established the pervasiveness of interpersonal conflicts and online harassment on social media, particularly for users from marginalized communities~\cite{pewsurvey, thomas2021sok, ashktorab2016designing, levy2022understanding}.
Many forms of online harassment involve the misuse of private information, partly because nearly half of all victims reported knowing their harasser, including acquaintances, family members, and ex-romantic partners~\cite{pewsurveyprivacy, aliapoulios2021large}. 
Examples of such incidents range from doxxing, where personally identifiable information is publicly disclosed~\cite{snyder2017fifteen, douglas2016doxing}, to non-consensual intimate imagery, where intimate photos are shared without permission~\cite{henry2020image, batool2024expanding}.
To address such harms, platforms and communities have developed a range of content moderation mechanisms~\cite{seering2019designing, jhaver2018online, crawford2016flag}.
In the following, we outline key dimensions of these moderation mechanisms, with a particular focus on how user reporting systems fit within this landscape.}

\subsubsection{\textbf{User reporting versus automated detection}}
\revise{Some approaches aim to prevent harm before it occurs—for example, by deploying nudges that warn users before posting potentially toxic content~\cite{ashktorab2016designing, seering2019designing}.
Many others focus on surfacing abusive behavior shortly after it happens.
For instance, many platforms and communities have adopted automated algorithms to detect abusive content (e.g., Facebook~\cite{FacebookAlgorithm}, Reddit~\cite{Chandrasekharan2019, jhaver2019human}, Discord~\cite{choi2023convex}).
In addition, many also maintain reporting systems that allow users to flag problematic content for moderator review~\cite{crawford2016flag, kou2021flag, wang2023reporting}.
While automated systems can scale efficiently, they often struggle with detecting nuanced and complex incidents and adapting to diverse community norms~\cite{kumar2021designing, seering2020reconsidering}.
Moreover, automated moderation is often inappropriate in private spaces where users have heightened privacy expectations, such as E2EE messaging channels~\cite{kamara2022outside, pfefferkorn2022content}, voice chats~\cite{jiang2019moderation}, ephemeral messages~\cite{kotfila2014message}, or direct messages between individual users, as they constitute continuous monitoring of these private conversations.
\textit{In these private spaces, user reporting remains the primary mechanism for surfacing harmful behavior~\cite{pfefferkorn2022content, kamara2022outside}.}}

\subsubsection{\textbf{Automated versus manual review}}
\revise{Once potentially abusive content is detected, how it is handled can also vary widely, ranging from fully automated enforcement to manual review by human moderators.
On one end of this spectrum, automated systems may determine whether flagged content should be immediately removed or hidden~\cite{horta2023automated, FacebookAlgorithm, gorwa2020algorithmic}.
User reports may also be used as inputs to automated systems.
For example, on the E2EE platform Signal, a ``humanity check'' is triggered after multiple reports against an account~\cite{signalreporting}.
On the other end, human moderators continue to play an important role in reviewing flagged content and making final decisions~\cite{roberts2016commercial, FacebookLabor}, although automated systems may assist in triaging content or offering recommendations~\cite{halfaker2020ores, koshy2024venire}. 
\textit{While automation can be effective in addressing clear-cut cases, our work focuses on how human moderators investigate more complex user reports.}}

\subsubsection{\textbf{Platform versus community-level moderation}}
\revise{Content moderation systems also differ significantly between platforms and communities.
Here, \textit{communities} refer to groups of chat rooms within a larger platform, such as Slack workspaces~\cite{SlackWorkspace}, Discord servers~\cite{DiscordServer}, or WhatsApp communities~\cite{WhatsAppCommunity}.
First, platform and community moderation differ in the types of policies they enforce. Platform moderation typically focuses on violations of legal or platform-wide rules, such as copyright infringement, violence, or child abuse~\cite{gorwa2020algorithmic}.
In contrast, community-level moderation focus more on community norms, ranging from general ones shared across many communities (e.g., prohibitions against harassment) to highly specific rules unique to each community~\cite{fiesler2018reddit, chandrasekharan2018internet}.
Additionally, platform and community moderators differ in their available resources.
Platforms can develop and deploy sophisticated automated moderation algorithms~\cite{FacebookAlgorithm} and hire professional moderators to review content at scale~\cite{ruckenstein2020re}.
By comparison, community moderation is more constrained~\cite{seering2020reconsidering}.
Automated tools are often limited to simple keyword filters~\cite{jhaver2019human}, and moderation is carried out by volunteers or part-time community members~\cite{seering2019moderator}.}

\revise{These differences extend to their reporting systems.
Reporting systems of platforms are typically built into the platform affordances and follows a standardized, streamlined workflow~\cite{cover2025reporting}.
In contrast, community-level reporting tends to be more ad hoc, for instance, receiving reports via direct messages, moderation bots, or emails~\cite{reportDiscord}.
As a result, community moderators tend to have broad discretion in how reports are reviewed and addressed~\cite{cai2021after, jiang2019moderation}.
Besides, since community moderators are embedded in their communities, they can draw on personal familiarity with involved users and consult them in the decision-making process~\cite{jiang2019moderation, cai2021after}.
But it can also introduce bias, when moderators have social ties to those involved or are directly implicated in the conflict~\cite{liao2025building}.
\textit{In this work, we examine how community moderators leverage their discretion in the reporting process and explore design opportunities to support a more accountable and fair system.}}


\subsection{Research Efforts to Improve Reporting Systems}
Despite their intended role in addressing online harassment and conflicts, reporting systems are often perceived by users as ineffective, confusing, and time-consuming~\cite{kou2021flag, crawford2016flag, wang2023reporting, aliapoulios2021large}. 
Reflecting this frustration, 47\% of people who have experienced harassment choose not to report their harasser~\cite{pewsurvey}. 
Instead, many prefer to block or mute them, which unfortunately limits moderation systems’ ability to prevent bad actors from further harming other users~\cite{jhaver2018online, responding2021thorn}.
In more extreme cases, victims may resort to publicly exposing and shaming their harassers~\cite{sultana2021unmochon}. 

In response, a growing body of research seeks to improve online reporting systems, with particular focus on supporting evidence collection throughout the reporting process. 
Some tools help victims gather trustworthy records of ongoing harassment~\cite{goyal2022you, sultana2021unmochon}, while others assist moderators in profiling victims and harassers to inform their decisions~\cite{cai2021after, im2020synthesized}. Researchers have also documented how moderators use bots to track ephemeral messages~\cite{kiene2019technological}.
However, most of these tools address isolated parts of the evidence collection process. \textit{In this work, we aim to characterize the broader design space to inspire future research.}

Noting the similarities between online reporting systems and offline criminal justice, researchers have begun applying offline legal frameworks to redesigning online reporting systems~\cite{fan2020digital, im2021yes, xiao2023addressing}. 
Offline justice systems have long faced public criticism for being overly complex, slow, unfair, and lacking respect for victims and the broader public~\cite{law1999review}. 
Legal scholars and practitioners have thus proposed alternative frameworks to reform justice systems, including restorative justice~\cite{braithwaite2002restorative}, therapeutic jurisprudence~\cite{daicoff2006law, wexler1992putting}, and problem-solving courts~\cite{freiberg2001problem}. 
Inspired by this shift, there is a growing interest among social computing researchers in applying the restorative justice theory to online settings.
In particular, researchers criticized the punitive nature of content moderation systems, as punishment alone is ineffective in deterring harassment and repairing the harm experienced by victims~\cite{hasinoff2020promise, schoenebeck2021drawing}.
Consequently, various restorative justice practices have been explored, such as video-offender conferencing~\cite{xiao2023addressing}, and apology~\cite{ngoc2025design, schoenebeck2021drawing}.
\textit{While these approaches focus on post-adjudication interventions like punishment and rehabilitation, our work focuses on the adjudication process itself, i.e., how decisions are made in the first place.}

Additionally, online dispute resolution systems represent another line of research that adapts offline trials to resolving conflicts between online users~\cite{rule2016designing, aouidef2021decentralized, kou2014governance}.
For example, eBay once implemented community courts, where buyers and sellers submitted evidence in turn, and crowdsourced jurors reviewed the evidence to reach a final decision~\cite{rule2010leveraging}. 
Similarly, Wikipedia established a formal deliberation process for resolving content and policy disputes among editors, inviting the broader community to discuss and deliver a resolution~\cite{im2018deliberation}.
While these processes can lead to more perceived legitimacy when executed well~\cite{fan2020digital, pan2022comparing}, they are often criticized for being resource-intensive and prone to stalemates~\cite {im2018deliberation}. 
They differ from reporting systems in that they set up a structured trial between opposing parties, whereas reporting systems typically task moderators with investigating the case and rarely involve the reported person~\cite{cai2021after, jiang2019moderation}.
Nevertheless, we draw on the experiences and challenges of these systems to inform the redesign of online reporting systems.

\subsection{Offline Inquisitorial and Adversarial Legal Frameworks}
In this paper, we draw inspiration from the legal concept of adversarial and inquisitorial models of justice systems. In an \textbf{inquisitorial model}, the judge appointed by higher authorities assumes an active role in investigating the facts, directing the collection of evidence, and shaping the case development~\cite{damaska1972evidentiary, damaska1986faces}.
In comparison, in an \textbf{adversarial model}, the opposing parties have the primary responsibility for defining the issues in dispute, investigating the facts, and advancing their cases, whereas the judge plays a neutral, passive role in ensuring procedural fairness~\cite{damaska1972evidentiary, damaska1986faces}. 
\revise{As a result, prior work suggests that adversarial models are often perceived as offering higher procedural justice~\cite{thibaut1975procedural, thibaut1978theory, sevier2014truth}, while inquisitorial models are more strongly oriented toward truth-finding~\cite{sevier2014truth, lind1973discovery, damaska1986faces}; however, both procedural justice and accuracy depend heavily on jurisdiction-specific rules and implementation practices~\cite{turner2019purposes, bamberger2011privacy}.}
Inquisitorial systems are widely adopted in continental European countries like Germany and France, whereas adversarial models are commonly used in English-speaking countries like the United States and the United Kingdom~\cite{ploscowe1935development}.
In reality, there exist no purely inquisitorial or adversarial legal systems; instead, legal systems typically blend elements of both models, representing a spectrum of power distributions between judges and the opposing parties~\cite{thibaut1973procedural, freiberg2011post}.

How a government organizes its justice system reflects its broader mode of governance. 
In the book \textit{The Faces of Justice and State Authority}~\cite{damaska1986faces}, Damaska identifies two key dimensions: the organization of government and the legitimate function of government.
\revisecomment{Strength the claim
}{When government authority is centralized and hierarchical~\citep[pp.~47--56]{damaska1986faces}, and the government takes an active role in implementing policies~\citep[pp.~154--168]{damaska1986faces}, legal systems tend to favor inquisitorial approaches, which grants the government more authority to advance its objectives.
In contrast, when governance is decentralized~\citep[pp.~57--66]{damaska1986faces} and the government limits its role to only resolving conflicts~\citep[pp.~109--135]{damaska1986faces}~\cite{gabrieli2022conflict}, an adversarial model is more likely to be preferred, as the government delegates more responsibility to individuals and primarily serves as a neutral dispute resolver.}
In this work, we examine the inquisitorial practices within current reporting systems and explore a new design space for incorporating more adversarial elements.


\section{Methods}
\subsection{Research Site: Discord}
\revise{Our work focuses on community-level reporting systems that addresses interpersonal conflict and harassment in private spaces. 
We select Discord as the primary research site for two reasons.
First, Discord hosts a large and diverse ecosystem of communities with active community moderation. 
As of December 2025, Discord reports 689 million registered users and 260 million monthly active users~\cite{discord2025resourcera}.
Users can create their own communities, known as servers, ranging from small groups of friends to large communities with hundreds of thousands of members~\cite{aquino2025discord}.
Each server establishes its own moderation systems that are primarily operated by volunteer community members~\cite{seering2019moderator}.
While Discord moderators manually patrol communities and adopt automated tools to detect abusive behaviors~\cite{choi2023convex, yoon2025s}, reporting systems still play an important role in surfacing issues that occur outside moderators’ direct view~\cite{jiang2019moderation}. 
Moderators can sanction users by removing messages, muting them, or issuing temporary or permanent bans.}

\revise{Second, Discord supports a wide range of private spaces across texts, audios, and videos, making it well suited for our purposes. 
Users can create private channels restricted to selected participants and do not grant moderators access~\cite{discord2025permissions}.
Users can also communicate with other community members through direct messages, which sit outside of each community and are thus inaccessible to community moderators~\cite{discord2025blockingprivacy}.
Within these private spaces, users can initiate voice or video calls that are not automatically recorded by the platform, although participants may manually record them~\cite{jiang2019moderation}.
While Discord does not offer sophisticated ephemeral controls for text messaging such as self-destructing or view-once messages, users can edit or delete sent messages, with only an ``edited'' label visible and no accessible message history~\cite{discord2025messageapi}.
In addition, Discord has gradually introduced E2EE for audio and video calls so that these conversations become encrypted to moderators and even the platform~\cite{discord2025e2eeav}.
But text messages on Discord are not end-to-end encrypted. Because E2EE text messages are an important example of private spaces, we occasionally reference practices from E2EE messaging platforms such as WhatsApp and Matrix to complement our analysis.}

\subsection{Interviews}
To understand how reporting systems operate in Discord communities, we conducted semi-structured interviews with 5 community moderators and 7 users who had experience with reporting or being reported on Discord. We denote moderator participants as M1–M5 and user participants as U1–U7. We recruited moderators by directly messaging them on Discord; with moderators’ permission, we also posted recruitment messages in their servers to recruit users.
All interviews were conducted via video calls. Moderator interviews lasted 93 minutes on average, while user interviews lasted 47 minutes on average. Participants received a \$20 gift card as compensation.
Additionally, we also drew on closely related prior work that present rich details of community-level reporting systems on Discord and similar platforms~\cite{jiang2019moderation, cai2021after, seering2019moderator}.

\subsection{Data Analysis}
\revise{Our analysis proceeded in two stages, reflecting the two contributions of this paper.
First, we conducted an interpretive analysis informed by adversarial legal theory~\cite{charmaz2006constructing}. 
In HCI, interpretive analysis is used to apply an existing theoretical framework to examine and reinterpret empirical practices.
Our approach follows prior work that adopts this method to study sociotechnical systems through lenses such as affirmative consent~\cite{im2021yes} and restorative justice~\cite{xiao2023addressing}. We note that this use of interpretive analysis is distinct from \textit{interpretative phenomenological analysis}, which focuses on understanding how individuals make sense of their lived experiences~\cite{walsham2006doing, soden2024evaluating}.
We analyzed interview transcripts alongside curated literature on community-level reporting systems through this theoretical lens. We began with an initial round of open coding~\cite{saldana2021coding}, applying short, descriptive phrases line by line to keep codes close to participants’ accounts (e.g., ``initiate the report'', ``biased moderators'', and ``collect reported conversations'').
We then conducted interpretive analysis guided by key concepts from adversarial legal literature, including ``inquisitorial models'', ``documentary evidence'', and ``procedural justice'' (See Table~\ref{framework} for more examples).
Finally, we organized these codes into higher-level themes to arrive at our findings (See Table~\ref{thematic_analysis} for examples). 
The first and second authors independently conducted the interpretive analysis and met weekly with the research group to compare interpretations, refine themes, and iteratively revise the coding framework.}

\revise{Second, building on this interpretive analysis, we adopted threat modeling as an analytic method~\cite{xiong2019threat} to map out concrete reporting designs that incorporate adversarial elements.
Threat models helped us specify how increased user control over evidence collection could be implemented while constraining opportunities for misuse.
We describe our threat models in the Section~\ref{sec:evidence-collection}.}


\section{Rethinking Community-Level Reporting Systems through Adversarial Legal Frameworks}\label{sec:framework}

\subsection{Inquisitorial Practices of Community-Level Reporting Systems}

We argue that community-level reporting systems predominantly follow an inquisitorial model, because moderators hold primary authority over how evidence is collected and evaluated and tend to prioritize truth-seeking over the preferences or concerns of involved parties. In this subsection, we will examine the reporting systems of Discord communities through the lens of adversarial models and highlight the inquisitorial practices they embed.

\revise{It is important to clarify that not all moderators exercise their authority in strongly inquisitorial ways, nor do they do so in all situations. 
Our concern lies not with individual actions but with the lack of a structured process for engaging users’ concerns, but instead whether and how to involve users is often left to each moderator’s discretion.
We view the inquisitorial nature of community-level reporting systems as emerging from broader structural conditions.}
First, online reporting systems are shaped by the autocratic governance structure prevalent in online communities~\cite{zhang2020policykit, schneider2022admins}, where community administrators and moderators have broad privileges over regular users and actively enforce community norms~\cite{seering2019moderator}.
Second, most community moderators are volunteers or part-time contributors who typically lack formal training in conflict resolution~\cite{seering2019moderator}. 
Reporting systems in these communities are typically informal and loosely structured~\cite{cai2021after, jiang2019moderation}.
For instance, while some communities in our interviews adopted a ticket bot to slightly organize reporting procedures, others relied on ad hoc practices such as asking users to directly message moderators. 
Consequently, moderators tend to default to inquisitorial approaches to evidence collection.

\subsubsection{\textbf{Collecting Documentary Evidence}}
\revise{Documentary evidence refers to social media posts and conversation logs that capture the reported behavior and its surrounding context~\cite{rychlak1996documentary}.
Such evidence is often critical for moderation decisions~\cite{cai2021after, jiang2019moderation}. 
Community moderators typically take primary responsibility for gathering this evidence so that they can rely less on evidence submitted by users.
For example, M1 described the reporting workflow in their community:
``\textit{If a user wants to report someone, they open a ticket at [the name of their ticket bot] and answer several questions, like why they are reporting this person and which message caused the issue... It is then usually the moderators’ responsibility to find the evidence and decide what to do with them... If we do ever reach out to the reporting user, it’s usually only when evidence was deleted or when it happened through DMs.}''
When the reported incident happened in private spaces, moderator participants tend to be more skeptical, either asking for more contexts or simply ignoring such evidence.
For instance, M2 said ``\textit{I usually ignore the messages that have the edited tag}'', even when edits might be minor, such as fixing a typo.
For incidents in direct messages, M4 explained: ``\textit{We usually asked for screenshots from both parties. If both parties are unwilling to provide screenshots, we usually have to not moderate either}''.
}

\move{Move from Design Space}{When moderators contact other bystanders or the reported user directly for verification, they sometimes do not obtain the reporting user’s consent.
For instance, M2 recalled: ``\textit{In one of the cases that I remember, I don’t believe that I got the consent of the person reporting when contacting the reported person. This was probably because I needed to make a fast decision,}'' while also acknowledging that ``\textit{in other cases, getting consent from the reporting user would be a lot better.}''
Similarly, when verifying a reported incident in voice chats, M1 described reaching out to ``\textit{as many witnesses as we can at one time to understand how an event unfolded.}'' but did not take the time to get consent from the reporting person.}

\revise{This inquisitorial orientation is further reinforced by technical infrastructure adopted by community moderators.
For instance, many Discord communities deploy bots that automatically log edited or deleted messages into dedicated moderation channels to preserve evidence~\cite{jiang2019moderation}.
While such practices support moderators’ ability to investigate incidents, they can also introduce unintended privacy risks. 
As M4 described, ``\textit{We had a situation once where somebody sent a Nitro gift link by accident and deleted the message. But we could still see it in the logs, and they asked us, ‘could you please delete that from your logs?’}''
In these cases, users must explicitly request the removal of sensitive information that they may have assumed was ephemeral.}

\subsubsection{\textbf{Collecting User Identifiers}}
\move{Move from Design Space}{Many community-level reporting systems do not support anonymous reporting.
All moderator participants have access to the identities of reporting users, even when those users wish to remain anonymous.
As these moderators actively monitor community discussions, knowing who submitted a report enabled them to draw on contextual knowledge about the user (e.g., their personalities) or the reported argument~\cite{cai2021after}.
M2 recalled such an incident, ``\textit{I think it's actually really important not to make them anonymous. Let's say two users are having an argument and then they report each other for breaking rules. We can look at their history at the server to see if they've been problematic in the past or if they have a bias anyway.}''
On the other hand, many users actually want anonymous reporting. 
Prior work and our interviews both suggest that reporting users are worried that moderators might leak their account identifiers to reported users and other community members, which might trigger further harassment or victim-blaming~\cite{sultana2021unmochon, take2024stoking}.
Additionally, if moderators have a negative impression towards one of the opposing parties, knowing users' identities might bias their decisions.
As U7 mentioned, ``\textit{It really depends on whether there’s a conflict of interest with the moderator. For example, if the moderator is closer to the person I’m reporting, or if the moderator doesn’t get along with me, then I’d also want to stay anonymous.}'' }


\subsubsection{\textbf{Collecting Account Information}}
\move{Move from Design Space}{Moderators in most reporting systems rely on account information to inform their decisions, including users’ reporting and reported history, join date~\cite{cai2021after}.
Users generally recognize the value of this information for moderation purposes and do not object to its use~\cite{wang2023reporting}.
Some participants noticed that their Discord communities might collect additional demographic information (e.g., gender or country of residence) during onboarding, and expressed concern that such account information could bias moderators’ judgments.
As U1 articulated, ``\textit{Once a moderator knows that I’m a woman, he’ll think that my report is just me being oversensitive. It’s also about the IP address—some people think that just because I’m at a certain IP, I must be acting a certain way.}''}

\subsubsection{\textbf{Reporting Workflows}}
\move{Move from Discussion}{Inquisitorial tendencies also appear in how reporting workflows are structured.
Our interviews indicate that users typically have little control over which moderator reviews their case, even when they have concerns about bias.
As M5 recalled, ``\textit{There was a user that was reporting the behavior or someone...They said I would rather not have this moderator be here or be the one giving the punishment or anything, because I know they are biased against me.}''
In practice, moderator assignment is opaque, and excluding specific moderators is often difficult. 
M1 explained, ``\textit{If they reported through our [ticket] bot, they don't know who they're talking to—everyone in our team has access and can handle it.}''
After moderators make their decisions, users typically receive only a generic notification about the outcome, without knowledge of the underlying evidence or rationale, echoing findings from prior work~\cite{jhaver2019does}.
M1 described as such: ``\textit{Yeah, we don't refer to specific messages [in the notification]. We just tell them that they have broken this rule.}''
As such, users must proactively appeal to get more information, as mentioned by M2, ``\textit{Usually we make the decision to moderate them without their knowledge... If they have an issue with that, they open a ticket or ping a moderator to find out.}''}

\begin{figure*}
    \centering
    \includegraphics[width=\textwidth]{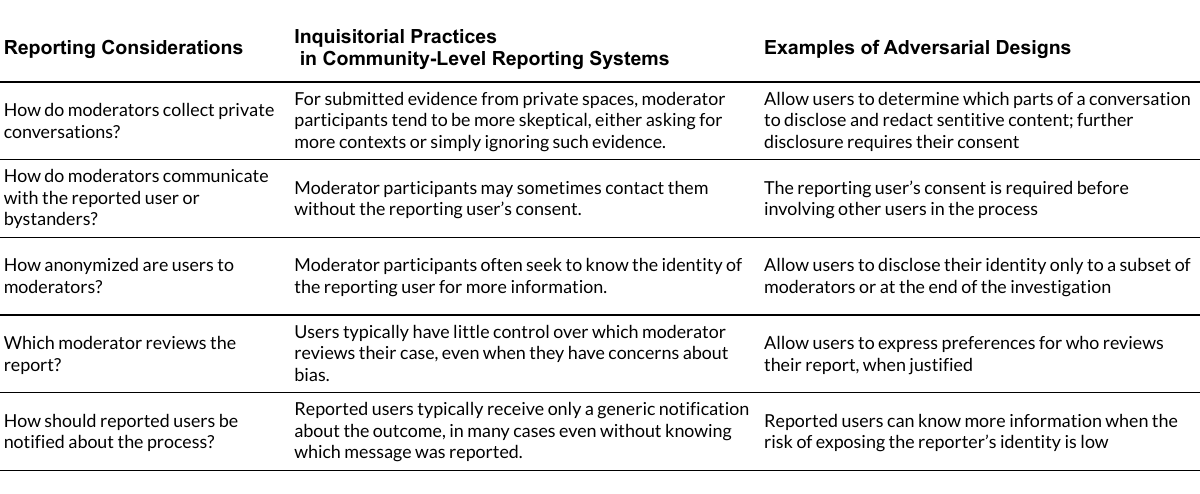}
    \caption{\revise{Drawing on interviews with moderators and users, we examine how community-level reporting systems often exhibit inquisitorial practices, where moderators take primary responsibility for gathering evidence and shaping cases. While not all moderators act this way in every situation, we highlight the absence of structured processes for consistently engaging users’ concerns, leaving such decisions to discretion of individual moderators. We contrast these observed practices with adversarial approaches, which grants users greater control while incorporating safeguards to prevent abuse.}}
    \hfill
    \label{framework}
\end{figure*}

\subsection{Weaknesses of Inquisitorial Reporting Systems Addressed by Adversarial Approaches}

Legal literature has offered extensive discussion about the strengths and weaknesses of adversarial and inquisitorial models~\cite{damaska1972evidentiary, damaska1986faces, kim2014adversarial, findley2011adversarial, ambos2003international}. 
Drawing on this literature, we examine how the inquisitorial nature of reporting systems can give rise to user frustrations in some cases.
\revise{Note that we do not argue that inquisitorial practices are always harmful. 
Indeed, they may align well with norms of some communities, for example when moderators are widely trusted or when risks of retaliation are low. 
Rather, our goal is to highlight scenarios in which inquisitorial approaches can lead to unintended harms, and to show how adversarial legal frameworks provide a useful lens for reasoning about alternative procedural designs that mitigate these harms.}

\subsubsection{\textbf{Procedural Justice}}
\textit{Procedural justice} refers to the fairness of the processes through which authorities make decisions and exercise power, as opposed to \textit{substantive justice}, which concerns the fairness of the outcomes themselves~\cite{tyler2006people}.
\revise{In this work, we adopt a \textit{normative} perspective on procedural justice, which highlights principles that are valued regardless of the case outcome, such as impartiality, neutrality, and respect for individual rights~\cite{tyler1988procedural, tyler2006people}.
This contrasts with the instrumental perspective, which holds that people’s sense of fairness is shaped primarily by whether outcomes favor them and how much influence they believe they have over the outcome~\cite{sevier2014truth}.}
Prior work shows that such perceptions of procedural fairness play a key role in shaping views of institutional legitimacy, which in turn affect whether individuals accept decisions and comply with rules or laws~\cite{tyler2003procedural, vsifrer2014assessing, hough2013legitimacy}.

\revise{Legal scholars have produced a rich body of work comparing perceptions of procedural justice in inquisitorial and adversarial legal systems.
Several studies suggest that adversarial models are often perceived as more just, as they better accommodate users' concerns and voices during the legal process~\cite{tyler2006people, thibaut1975procedural, houlden1978preference, burdziej2019fairness}.
At the same time, other research highlights that perceptions of procedural justice are shaped not only by legal structure but also by cultural expectations and how these systems are implemented in practice~\cite{anderson2003perceptions, brekke1991juries}.
Our interviews, along with prior research, reveal persistent user frustrations about the lack of procedural justice in reporting systems, particularly around limited visibility into what evidence is gathered and little opportunity to shape the process~\cite{xiao2023addressing, jiang2019moderation, wang2023reporting}.
These concerns reflect the inquisitorial structure of most community-level reporting systems, where moderators have broad discretion over what evidence is considered, how investigations proceed, and whether users are involved.
In response, we see promising opportunities to introduce adversarial practices that give users a more active role and, in turn, promote greater procedural fairness.}

\subsubsection{\textbf{Privacy Protections}}
Our analysis above shows how inquisitorial practices in community-level reporting systems can sometimes overlook users’ privacy concerns. For example, these systems may collect users’ identities, contact other involved parties without explicit consent, or encourage broad evidence collection. Such practices can harm users’ privacy and may also expose reporting users to further harassment or victim-blaming when their personal information is shared with hostile community members~\cite{take2024stoking}.

By contrast, adversarial systems may offer stronger privacy protections as a direct implication of enhanced procedural justice, since each side can have more say in evidence collection and presentation~\cite{damaska1972evidentiary}.
With their greater control over evidence collection, a party can choose to withhold information that might compromise their privacy if its value as evidence does not outweigh its risks.
Even when the opposing party seeks to introduce such evidence, rules in adversarial jurisdictions allow for its exclusion when requested, even if it is relevant~\cite{stacy1991search, findley2011adversarial}.
\revise{Importantly, we do not claim that actual inquisitorial legal systems in offline settings protect privacy less than adversarial ones. 
In practice, privacy protections depend on many factors, including jurisdiction-specific rules, culture values, and implementation practices~\cite{turner2019purposes, bamberger2011privacy}. 
Our claim is more limited: at a conceptual level, adversarial models emphasize participant control and contestation in ways that can inform the design of reporting systems that are more responsive to users’ privacy concerns.}

\subsubsection{\textbf{Accuracy of Decisions}}
Finally, there are mixed findings regarding whether inquisitorial models yield more accurate decisions than adversarial ones~\cite{sevier2014truth, lind1973discovery, thibaut1972adversary}. 
In inquisitorial models, moderators, as third parties to the reported conflict, can collect evidence without bias in theory.
However, legal research has shown that judges tend to be more diligent when the case involves the interests of their higher authorities compared to when it concerns private interests~\cite[p.54]{damaska1986faces}. 
These concerns are amplified in online settings.
Community moderators often lack the time, training, or incentive to conduct thorough investigations, as many serve in part-time or volunteer roles~\cite{seering2019moderator}.
This becomes more problematic as online harassment grows more persistent and sophisticated, demanding more time and effort from moderators~\cite{pewsurvey}. 

In contrast, the competitive nature of the adversarial process creates stronger incentives for each party to thoroughly search for evidence than moderators typically have in inquisitorial settings~\cite{kim2014adversarial}.
While this can lead to a richer collection of evidence, it also raises risks: parties may be tempted to manipulate or selectively present evidence to favor themselves~\cite{lind1973discovery}. 
\revise{To obtain an accurate final decisions in adversarial models also requires that parties have equal access to tools and resources for gathering and authenticating evidence, highlighting an important opportunity for design interventions~\cite{findley2011adversarial}.}


\subsection{Considerations for Incorporating Adversarial Practices into Online Reporting Systems}

\subsubsection{Degrees of Inquisitorial versus Adversarial Practices}
Offline jurisdictions have recognized that purely adversarial or purely inquisitorial systems can cause more harm than benefit, and have thus begun adopting hybrid models.
For example, inquisitorial elements have been integrated into American systems to reduce costs and delays~\cite{findley2011adversarial}, while adversarial practices have been introduced in Continental Europe to enhance perceptions of procedural justice~\cite{walpin2003america, van2003adversarial}. 
Therefore, we envision that online reporting systems can similarly benefit from incorporating adversarial elements.

\move{Move from Discussion}{Importantly, incorporating adversarial practices does not imply adopting adversarial procedures wholesale, as fully adversarial procedures pose significant practical challenges in online community settings.
First, because community administrators and moderators are granted broad authority over regular users \cite{zhang2020policykit, schneider2022admins}, they may be reluctant to adopt reporting processes that substantially reduce their direct control over investigations.
Second, community moderators often operate under tight time constraints and without formal training~\cite{seering2019moderator, yoon2025s}, which makes fully adversarial processes difficult to sustain in practice, as they typically require greater effort and oversight~\cite{findley2011adversarial}.}
For these reasons, we argue that \textbf{the best approach is a hybrid one that incorporates some adversarial elements into the current inquisitorial approach}.
This also motivates our focus on identifying technical affordances that embed adversarial practices within existing inquisitorial workflows, thereby reducing the practical burden on community moderators of learning and managing adversarial processes.

\subsubsection{Threats of Abuse in Adversarial Reporting Designs}
A central threat with incorporating adversarial elements is that with more control over the process, users are more likely to abuse that control for self-interest~\cite{damaska1972evidentiary, kim2014adversarial}.
For example, if users can influence who reviews their report, they may prefer moderators who are more sympathetic to their perspective. 
And while adversarial models encourage users to actively gather evidence, they also increase the risk of evidence manipulation or selective disclosure~\cite{lind1973discovery}. 
Such abuse can be even more prevalent in online settings where false reports are already common, whether to overwhelm moderators~\cite{matias2015reporting}, or frame other users~\cite{kou2021flag}.

\revise{Abuse risks also vary across users and communities with different technical expertise and norms. 
Many users are constrained by platform affordances and may engage in relatively low-tech forms of abuse, such as sending abusive content only in ephemeral conversations~\cite{jiang2019moderation} or submitting edited screenshots of conversation as evidence. 
In contrast, technically sophisticated users may exploit deeper system vulnerabilities, for example by manipulating clients or networks, or by leveraging compromised accounts to evade detection~\cite{munyendo2024you,doctorow2020eff}.
Additionally, abuse risks are not limited to reporting or reported users.
For instance, reporting users may also worry that moderators could misuse their discretionary power, for instance by favoring certain users or mishandling sensitive information~\cite{wang2023reporting}.}
Taken together, these considerations highlight why \textbf{threat modeling is crucial when coming up with adversarial practices that can guard against abuse}. Technical affordances that support adversarial practices must be designed with a clear understanding of who might abuse the system and how they might do so.

\subsubsection{Strategies to Prevent Abuse in Adversarial Reporting Designs}
To prevent potential abuse, the primary strategy offline adversarial jurisdictions employ is to incorporate the perspective of the opposing party.
For instance, the opposing party can cross-examine submitted evidence to test its authenticity~\cite{schlesinger1963elements} or file motions to challenge biased judges.
Such an approach is feasible in offline settings because defendants must be formally notified and allowed to participate in the process~\cite{procedural2025justia}.
To prevent retaliation, courts can issue protective orders or impose severe penalties for retaliatory behavior~\cite{retaliating2025cornell}. 
However, online communities are less capable of doing so as their punishments are often less effective and can be easily evaded, for example, by creating a new account or moving to a new platform.
As M3 recalled, ``\textit{We've had a case where two users who had a very personal vendetta against each other. Both of them made around 100 different alt accounts to simply join the server to spam or say bad things about the other person.}''
Consequently, many platforms choose not to engage the reported user during investigations, making practices like cross-examination infeasible~\cite{report2025messenger}.

Given these major differences in offline versus online systems, we choose in this work to \textbf{not rely on involving the opposing party in the reporting process to address all abuse but instead explore the design space of techniques that strengthen user-provided evidence}.
In particular, we draw inspiration from various rules and procedures used in offline adversarial systems to prevent abuse.
For instance, submitted evidence is evaluated against strict admissibility rules to ensure that only credible evidence is considered at trial~\cite{damaska1972evidentiary}. 
While online reporting systems could still involve the reported person during the investigation, we leave consideration of these possibilities to future work.
 



\section{Mapping the Design Space of Evidence Collection using an Adversarial Model}\label{sec:evidence-collection}
\revise{Our analysis motivates the exploration of evidence collection techniques that can be effectively integrated as adversarial practices within existing reporting systems.
We focus specifically on documentary evidence, which remains central to moderation decisions~\cite{cai2021after, jiang2019moderation} but has become increasingly difficult to collect due to emerging technical affordances in private spaces (e.g., voice channels~\cite{jiang2019moderation}, end-to-end encrypted chats~\cite{wang2023reporting}).}
In what follows, we identify two key challenges in granting users greater control over collecting and presenting documentary evidence in the reporting process.
Using threat modeling~\cite{xiong2019threat}, we analyze potential risks of abuse by different actors and examine trade-offs among competing design choices.
We then propose new designs that aim to support both user agency and the integrity of submitted evidence.

\textbf{Threat Models.}
A central question our designs must address is how to grant users more control in the reporting process while preventing abuse of that control.
We consider a threat model in which (1) platforms are trusted to see user messages and operate without bias, and (2) users and moderators are user interface (UI) bounded in their ability to abuse the system---that is, their actions are limited to what the platform's interface allows.
\revise{Our threat model considers the possibility that moderators may misuse the reporting system because they are often volunteer community members and deeply embedded in their communities~\cite{seering2019moderator, yoon2025s}.
In contrast, platforms tend to be less directly involved in community-level interpersonal conflicts~\cite{roberts2020behind}.
Moreover, most social media users, including community moderators, lack the technical expertise to circumvent platform interfaces~\cite{pew2019digitalknowledge}, making UI-bounded abuse a practical and realistic assumption in our context.}
Of course, it is possible for interactions to occur outside of this threat model: users can escape the UI through the use of modified applications~\cite{doctorow2020eff,munyendo2024you}, or users may not fully trust the platforms, especially those who intentionally choose E2EE platforms~\cite{ermoshina2016end}.
We will revisit these alternative threat models and discuss their implications for our proposed designs in Section~\ref{sec:discussion}.

\subsection{\textbf{Disclosure of Documentary Evidence}}

In adversarial models, users determine which conversation to disclose to moderators. 
But this poses threats to accuracy of moderation decisions: users might selectively present messages to favor themselves, or to frame the other party~\cite{lind1973discovery}. 
Users can simply justify such manipulation under the guise of privacy concerns. 
This reflects \textbf{an inherent trade-off between protecting user privacy and giving moderators sufficient information to make informed decisions in the process of evidence disclosure}.
This trade-off also mirrors the data minimization principle in the EU's General Data Protection Regulation (GDPR), which states that ``personal data shall be adequate, relevant, and limited to what is necessary in relation to the purposes for which they are processed''~\cite{regulation2016regulation}.
\textbf{To help navigate this tension, we draw inspiration from various data minimization practices to outline a design space that consists of four dimensions}~\cite{kulshrestha2021identifying, biega2020operationalizing, orekondy2018connecting}.
Communities can adopt the designs that best align with their privacy commitments.

\begin{figure*}[b!]
    \centering
    \includegraphics[width=\textwidth]{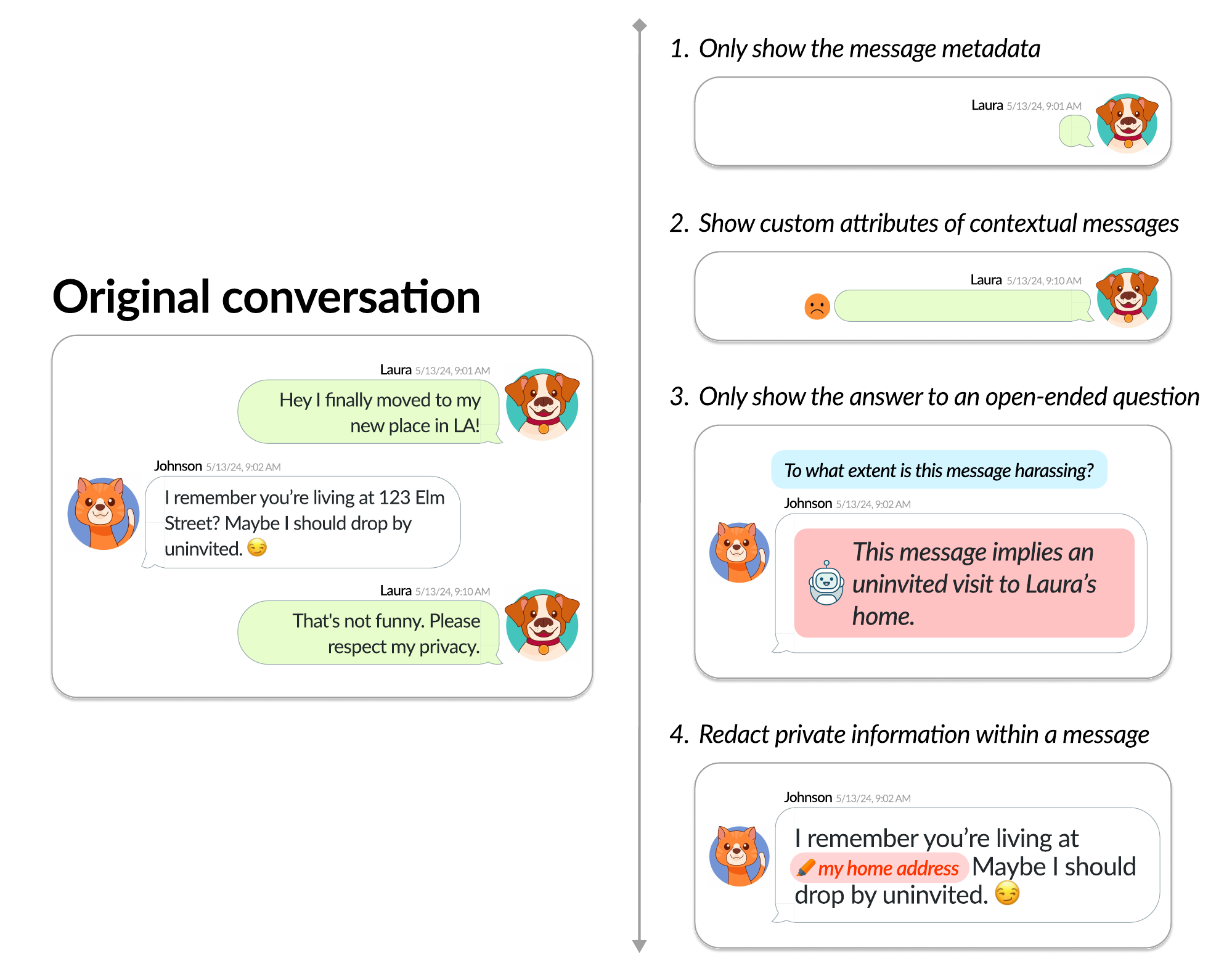}
    \caption{\textbf{Design Space for Information Visibility}. These design ideas illustrate varying levels of information visibility beyond fully hiding or fully revealing content. 
    (1) Only metadata like the sender and timestamp is shown. 
    (2) Custom attributes, such as sentiment and message length, are displayed. 
    (3) A language model answers an open-ended question, e.g., ``To what extent is this message harassing?'', while constrained to avoid referencing private details. (4) Users redact sensitive information, such as replacing their home address with the label ``my home address.''
    Although these examples focus on textual messages, the same ideas can be extended to images or videos.}
    \hfill
    \label{information visibility}
\end{figure*}

\subsubsection{\textbf{Design Dimension 1: Scope of Relevant Conversations}}
\revise{Before considering redacting private information within conversations, we should first determine the scope of relevant conversations.
If moderators have access to the relevant chat logs, they often read the full history to gain context, sometimes going far beyond what was initially flagged.
As M3 noted, ``\textit{despite what you said in the report, we need to go to the chat history and basically read from the start to get a full picture of what happened.}
Conversely, when moderators lack access to a private chat, they ask users to submit screenshots, giving them full control over what is included~\cite{report2025discord}.
These two extremes can result in either unnecessary privacy exposure or selective framing.
Instead, a more structured approach to defining the scope of relevant conversations can help avoid inconsistent and ad hoc decisions.
We also recognize that overly rigid scoping rules may fail to account for the variety of situations that arise in online conflicts.
To address this, reporting systems could offer a set of scoping options that balance contextual understanding with user privacy.}
\begin{itemize}
    \item Retrieving only \textit{the most recent messages in the reported conversation measured by count} can provide immediate context without overexposing unrelated content.

    \item When users have intense arguments with each other, it may be more appropriate to filter by \textit{all messages within a specific time window}.

    \item In a busy Discord channel, where several conversations often run together without distinct threads, filtering by \textit{messages from selected participants} can be more effective.

    \item These methods only focus on the specific reported conversation, but disputes may span multiple places, such as moving from public channels to direct messages. 
    In these settings, we might want to apply the aforementioned filters to all conversations between the involved users, not just the conversation where the report was initiated.
\end{itemize}


\subsubsection{\textbf{Design Dimension 2: Information Visibility}}
For each message within the scope of relevant conversations, the next step is to decide how much information to disclose. 
As illustrated in Figure \ref{information visibility}, this decision spans a spectrum, from fully hiding the message to revealing it in full.

\begin{itemize}
    \item 
    First, users can choose to completely remove a message. 
    While Discord does not currently support this capability, selective forwarding on platforms like WhatsApp \cite{whatsapp2017forward} and Signal \cite{signal2017forward} shows that omitting messages without leaving placeholders is a familiar interaction pattern.

    \item Second, users can blur the message content while still showing certain metadata like the timestamp and sender. This helps moderators understand messaging patterns in a conversation---for instance, when there is an overwhelming number of messages sent from one user. 

    \item Another option is to display custom attributes of a message, computed using predefined algorithms. Examples include its length, the presence of certain keywords/links/images, or even message sentiments. This approach has been explored in sensitive contexts like detecting child sexual abuse material~\cite{kulshrestha2021identifying}.

    \item With the growing capabilities of large language models (LLMs), including locally-hosted private models, we envision replacing predefined algorithms with more powerful LLMs. Instead of outputting a single attribute, these models could answer more targeted questions, for example, who initiated the dispute or how long the reporting user has experienced harassment. Nevertheless, this option is viable only if users trust the model, whether it runs locally on their device or through a trustworthy third‑party service.
    Otherwise, sending private conversations to the model only raises additional privacy risks~\cite{zhang2024s}. It is equally important to carefully design the prompts to avoid prompt injection attacks~\cite{liu2023prompt} and hallucination risks~\cite{huang2025survey}.
    Users should also have the right to review LLM answers before sending them to moderators.

    \item Users can redact private information within a message, by fully blurring it~\cite{orekondy2018connecting} or replacing it with a general term (e.g., replacing their exact home address with ``my home address'')~\cite{chow2009sanitization}. This is also familiar to reporting users on Discord: U7 would ``\textit{blur out [her] own picture and phone numbers, since sometimes when people send screenshots of text messages, both sides’ phone numbers get exposed in the chat log.}''

    \item At the most transparent end of the spectrum, moderators can view the full original content of the message, as in most current reporting systems. 
\end{itemize}

\begin{figure*}[t!]
    \centering
    \includegraphics[width=\textwidth]{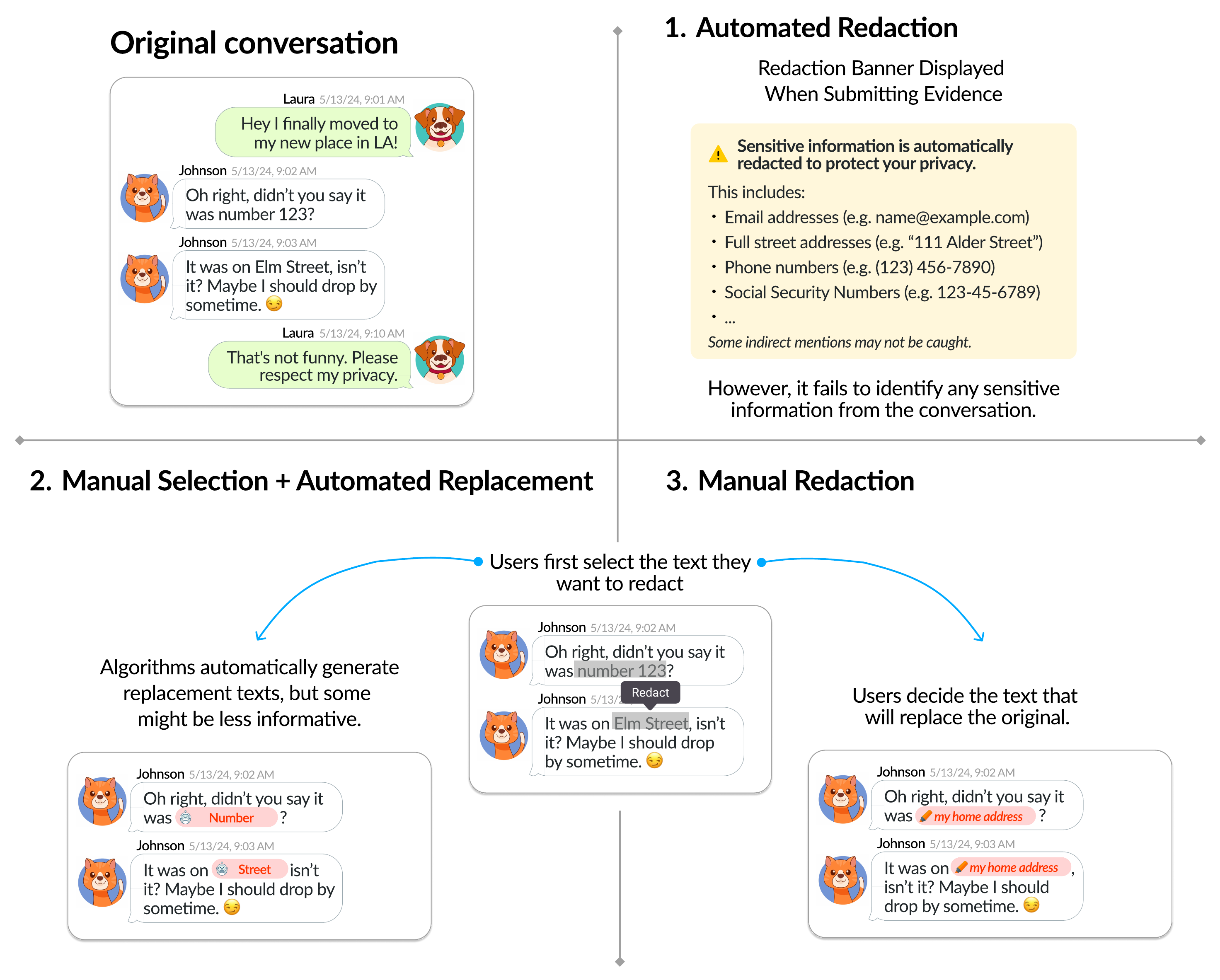}
    \caption{\textbf{This diagram illustrates how a redaction tool can support varying degrees of user control.}
    (1) \textbf{Automated Redaction}. An algorithm automatically detects and redacts sensitive information using a generic label. While this approach limits user manipulation, it may fail to catch all sensitive content. 
    (2) \textbf{Manual Selection + Automated Replacement}. Users manually select text to redact, and the system replaces it with a standard placeholder. This offers a balance between user input and automation, but the resulting redactions may be less informative for moderators.
    (3) \textbf{Manual Redaction}. Users have full control over redactions. This provides the strongest privacy protection but is vulnerable to misuse, as users may redact in ways that alter the intended meaning of the message.}
    \hfill
    \label{user_control}
\end{figure*}

\subsubsection{\textbf{Design Dimension 3: Access Control}}
After finalizing the documentary evidence, we should then determine who can access it. 
Most reporting systems grant access to all moderators and may archive all shared evidence for future audits by community admins.
For instance, M4 described that in his Discord community ``\textit{when we close the ticket, we would take a transcript of it, which is saved in another channel. And then we can see a transcript of everything that was said in that report}.''
From a threat modeling perspective, this creates potential threats to user privacy if access is overly broad or long-term.
Additional access control mechanisms can mitigate these risks.
For example, users may prefer to share their conversations only with a trusted subset of moderators.
Indeed, M5 noted that reporting users sometimes reach out to a specific moderator they trust, especially when the report involves \textit{someone on the staff team or someone important in the community}.'' In these cases, M5 described asking the reporting user, \textit{are you comfortable if I share this screenshot with the other mods?}'' before expanding access.
Access can also be limited to a specific time window or a fixed number of views. 
These restrictions help ensure that the evidence is used solely for moderation purposes and reduce threats from malicious community moderators.

\begin{figure*}[b!]
    \centering
    \includegraphics[width=\textwidth]{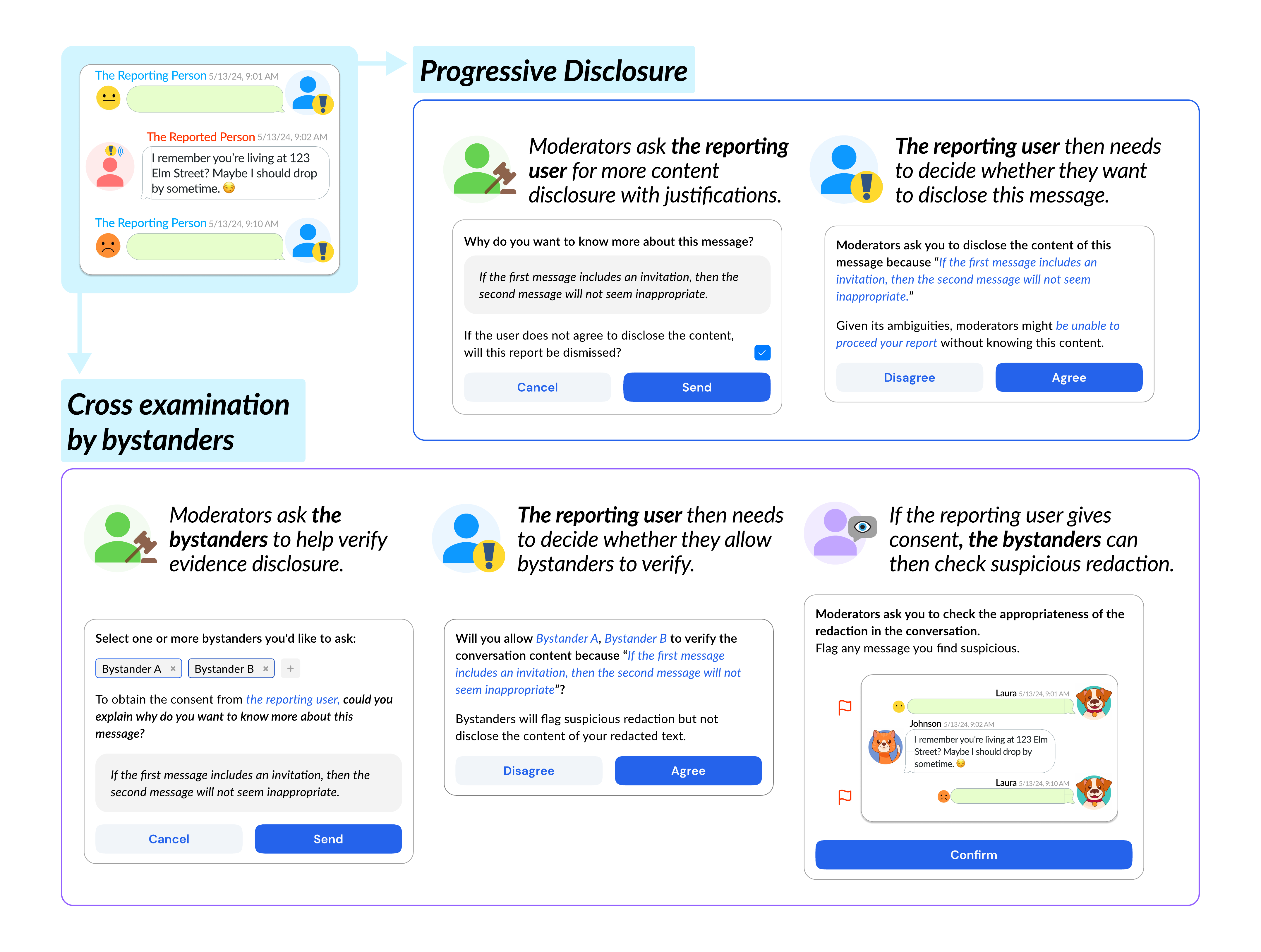}
    \caption{\textbf{Complementary Designs to Prevent Selective Disclosure}. Granting users more control in evidence disclosure increases the threats of selective disclosure by bad-faith reporting users. To prevent such abuse, we propose two complementary designs that involve the reporting person or bystanders. \textbf{Progressive Disclosure}. Moderators request the reporting user to reveal the content of a redacted message that is deemed critical for decision-making. In this request, moderators are expected to include justifications and explain the consequences of declining, which helps the reporting user better weigh their privacy against the goals of reporting.
    \textbf{Cross Examination by Bystanders}. Alternatively, moderators can also invite bystanders to help verify evidence disclosure from the reporting person. 
    However, moderators need to first obtain consent from the reporting person, who evaluates the privacy risks of involving bystanders in a reporting process. 
    In this example, bystanders can only flag potentially suspicious redactions but not disclose their content to moderators.}
    \hfill
    \label{progressive disclosure}
\end{figure*}

\subsubsection{\textbf{Design Dimension 4: User Control over Evidence Disclosure}}
A broader, high-level question remains: how much control should users have over evidence disclosure? Answers to this question map onto a spectrum between inquisitorial and adversarial models.
In a fully inquisitorial system, moderators---or automated systems reflecting moderators’ preferences---can access conversations freely without any user input.
In contrast, a fully adversarial model gives users complete control over what evidence is shared.
For each dimension described above, we still needs to answer who gets to decide on specific design choices---users or moderators.
And each specific design idea above can support varying degrees of user customization.

Take the dimension of conversation scope for example. 
A reporting system may automatically apply a particular filter, such as including the most recent 50 messages. 
To give users more control, a system might allow them to adjust the filter parameters (e.g., the number of messages) or choose from a list of predefined filters (e.g., time-based or participant-based).
At the highest level of control, users can freely submit any conversation.
Similarly, for information visibility, systems can vary in how much redaction control they give users, as illustrated in Figure \ref{user_control}.
If a reporting system allows users to redact private information, it can use predefined scripts to automatically redact sensitive fields such as home addresses or real names~\cite{gilbert2015open}. 
A more user-driven system might allow users to manually select what to redact, with the platform automatically replacing it with a generic placeholder.
Some systems can further allow users to specify their own replacement words.

Through the lens of our threat modeling, any form of user control over evidence disclosure carries the threat of selective disclosure in an attempt to mislead moderators.
For example, users might intentionally omit their own inappropriate messages to appear more benign to moderators.
Users might also move [their arguments] to direct messages simply to evade moderation, as M4 explained, ``\textit{Someone moved over to the direct messages, strictly for the purpose because they knew that [their behaviors] broke some server rules.}''
They might reveal only that a particular toxic keyword was used in a message from the opposing party, even if the message itself is not toxic in context.
Users might also withhold access from moderators who are familiar with the full context of the conversation.
As depicted in Figure \ref{progressive disclosure}, we propose two complementary designs that involve different other parties to mitigate such abuse.

\begin{itemize}
    \item First, we envision a model of progressive disclosure between moderators and users. Moderators can request access to specific portions of a minimized conversation and are expected to explain the rationale for their request. While users may choose to deny these requests, doing so may reduce the credibility of their evidence. If the withheld content is central to the case, the report itself may be dismissed.
    M4 adopted similar practices in his moderation, ``\textit{We usually asked for screenshots from the reporting users. If they are unwilling to provide screenshots, we usually have to not moderate either, because it's always innocent until proven guilty instead of guilty until proven innocent.}''

    \item Second, reporting systems could allow other conversation participants to verify whether evidence was disclosed appropriately. If there is sufficient justification to invite the reported person, this then bears some similarities to cross-examination in offline adversarial models. Their level of involvement can vary: they might provide a simple yes/no answer, flag suspicious messages, or disclose specific messages they believe were inappropriately excluded. With this additional information, moderators can then reach out to the reporting user for clarification or further disclosure.
\end{itemize}

\subsection{Authentication of Documentary Evidence for Persistent Conversations}
Even if users do not selectively disclose their conversations, any adversarial system must address the inherent communication gap between the informed users and the uninformed moderators~\cite{kim2014adversarial}.
\textbf{When users submit their evidence, they need to prove to moderators that their submitted conversations happened~\cite{messenger2017facebook}, known in offline settings as evidence authentication}.
While moderators may sometimes have enough trust in a user so that they accept the conversation as authentic~\cite{rajendran2024deniable}, the growing threats of false reports demands the authentication of user-submitted conversations~\cite{sultana2021unmochon}. 

In particular, authentication of online conversations consists of two key components: 1) \textit{identity attribution} that verifies the participants in the conversation are indeed the parties involved in the reported incident; 2) \textit{message authentication} that ensures that the messages in the conversation have not been altered and originate from the claimed sender.
When moderators can access the conversation directly, authentication is relatively straightforward, as in the case of community moderators reviewing messages in public channels.
However, community moderators often lack access to certain conversations, such as direct messages between users, conversations in private channels, or E2EE conversations.
These cases demand specialized authentication methods, which can be categorized based on who provides the authentication: the platform or the user who submits the evidence.
In the following, \textbf{we criticize the flawed use of these methods in existing reporting systems and then propose new design ideas that provide sufficient authentication}.

\begin{figure*}
    \centering
    \includegraphics[width=\textwidth]{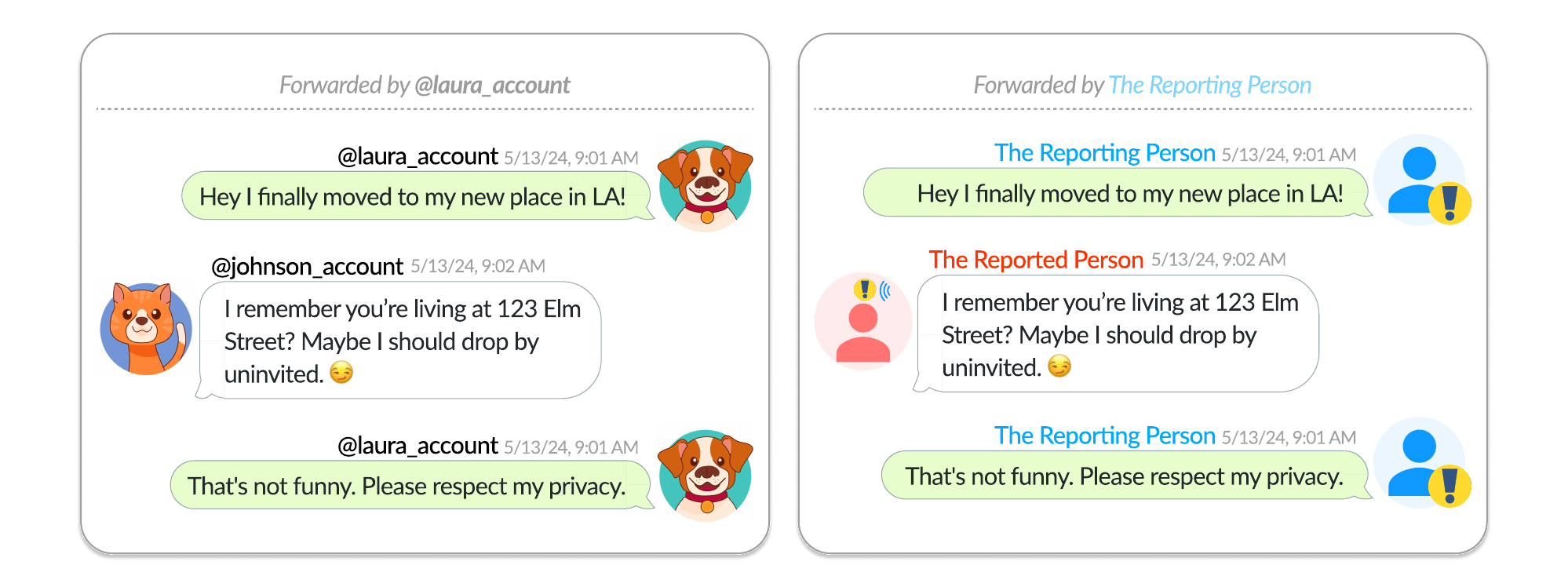}
    \caption{\textbf{Authentication by Platforms through Forwarding Messages from Moderators' Perspective}.  (Left) This example shows how message forwarding can be repurposed to support platform-side authentication. 
    The interface should indicate the message’s forwarded status and display account identifiers rather than usernames. 
    (Right) When anonymized reporting is supported, the interface should instead verify and label the roles of the reporting and reported users without revealing their identifiers.}
    \hfill
    \label{forwarded authentication}
\end{figure*}

\subsubsection{\textbf{Design Idea 1: Authenticated by Platforms}}
First, users can turn to platforms for authentication support.
On most non-E2EE platforms like Discord, textual conversations, including those in private channels or direct messages, remain accessible to the platform and can therefore be authenticated.
If Discord were to adopt end-to-end encryption (E2EE), however, the platform would no longer have default access to message content.
Nonetheless, cryptographic protocols such as message franking can still enable message authentication to community moderators in such settings ~\cite{grubbs2017message, tyagi2019asymmetric}.
For example, Facebook Messenger and WhatsApp incorporate such protocols to let users generate verifiable evidence of a reported message’s origin and integrity without exposing other unrelated messages~\cite{grubbs2017message, tyagi2019asymmetric}.

Nevertheless, platforms like Discord typically limit authentication support to their internal reporting systems, excluding community-level reporting systems.
Therefore, users must rely on third-party tools for authentication when reporting to community moderators~\cite{sultana2021unmochon, goyal2022you}. 
For instance, many platforms offer OAuth access to conversations~\cite{hwang2024adopting}, enabling third-party tools to authenticate message content~\cite{goyal2022you, rcfilter2017thomson}. 
These tools retrieve both content and metadata (e.g., author, timestamp) to verify message authenticity.
However, since OAuth access is typically granted to entire conversation histories rather than specific messages~\cite{discord2025oauth}, such platform-side authentication introduces privacy risks for reporting users.

We argue that online platforms should grant community moderators access to message authentication functionalities, given their central role in addressing online harassment~\cite{seering2020reconsidering}.
One promising direction is to appropriate existing message forwarding features, which are common across social media platforms and already familiar to users~\cite{yadav2023cryptographic}. 
However, current forwarding functionalities on nearly all platforms lack sufficient authentication support. 
For instance, WhatsApp does not show the author or timestamp in forwarded messages~\cite{forward2025whatsapp}, and Signal does not even mark messages as forwarded~\cite{forward2025signal}.
We argue that platforms should display additional information when a message is forwarded for reporting purposes.
Specifically, forwarded messages should clearly indicate their forwarded status, display the account identifier of the original author, and show the original timestamp (see Figure \ref{forwarded authentication}).

\subsubsection{\textbf{Design Idea 2: Authenticated by the User Who Submits the Evidence}}
Due to the absence of user-friendly authentication tools provided by platforms, users have to perform client-side authentication. 
The most widely used method is to take a screenshot of a conversation~\cite{sultana2021unmochon, yadav2023cryptographic}.
Unlike platform-side authentication, this approach increases the risk of abuse because users can easily tamper with screenshots or even alter what is displayed on their devices.
The former vulnerability is particularly concerning given easy access to image tampering tools~\cite{gupta2013faking, zheng2019survey}.
M3 mentioned such an incident: ``\textit{someone took a screenshot of their DMs and then edited it to make [the other person] look like doing something that broke the rules. We ended up taking a closer look at it, realizing that the font was slightly off and the messages were spaced incorrectly.}''
To address th concern of image tampering, some third-party tools streamline the screenshot process by sending the screenshot directly to moderators, reducing opportunities for manipulation~\cite{sultana2021unmochon}. 
Additionally, some communities require users to screen record their conversation interface, as M5 explained, ``\textit{it is a lot harder to fake for a recording of somebody scrolling through the message.}''

However, users can still manipulate content prior to capture—such as modifying HTML elements on desktop. To address this, moderators sometimes ask users to refresh the interface mid-recording, although moderators generally trust screen recording of mobile devices~\cite{yadav2023cryptographic}.
This layered approach mirrors the chain of custody in offline practices, where each step of evidence collection is documented to ensure its authenticity~\cite{giova2011improving}.

Unfortunately, these approaches only authenticate message content but fail to sufficiently verify identity. In other words, moderators do not verify that the participants in recorded images are truly the ones involved in the incident. 
Because most social media platforms display only a user's avatar and username instead of their full account identifier on conversation interfaces, malicious users can easily mimic these visual cues with a fake account to forge conversations.
Unmochon~\cite{sultana2021unmochon} partially addresses this by extracting account identifiers from Facebook Messenger’s URL when capturing screenshots of direct messages. However, this technique does not generalize to other platforms like Discord or to group chats, where URLs typically do not indicate participant identities.

When platforms do not take the initiative to implement such features, third-party tools can help enable client-side authentication. 
Web automation techniques can be used to collect evidence while preserving the chain of custody~\cite{pu2023dilogics, yeh2009sikuli}.
After users specify the conversation scope and involved parties, such tools could automatically record the screen, first scrolling through the selected conversation, then opening each user profile to display their account identifiers. By sending the resultant recording directly to moderators, we can then prevent users from tampering with it.

\subsection{Authentication of Documentary Evidence for Ephemeral Conversations}

The authentication approaches discussed above assume that messages are persisted, either on platform servers or on the client. 
Many reporting systems hold the same assumption, as reports are typically initiated by clicking on a specific message or the harassing user.
However, this assumption fails in ephemeral interactions---such as voice chats~\cite{jiang2019moderation}, virtual reality social media~\cite{freeman2022disturbing}, self-destructing messages~\cite{booth2021whatsapp}, or messages that can be edited or deleted after being sent~\cite{schnitzler2020exploring}. 
Although ephemeral messaging mirrors offline communication and can enhance privacy~\cite{kotfila2014message}, its transient nature is exploited by malicious actors to evade both message authentication and identity attribution~\cite{goyal2022you, kiene2019technological}.
By the time a user realizes they were harassed and decides to report it, the relevant context or even the harassing messages themselves may no longer be available. 
For instance, M1 mentioned that ``\textit{There was a point where someone was streaming incredibly gory NSFW things in the voice chat. We weren't able to like send mods there, and people in the chat weren't able to screen record it in time, before they stopped it.}''
Ephemeral conversations can also obscure harasser identities.
For example, in a voice channel with many simultaneous speakers~\cite{jiang2019moderation}, or in a crowded virtual reality game~\cite{blackwell2019harassment}, users may be unable to determine who initiated the harassment.
In the following, \textbf{we similarly criticize the shortcomings of two approaches used to authenticate ephemeral conversations, based on which we propose new design ideas that address such shortcomings}.

\begin{figure*}[b!]
    \centering
    \includegraphics[width=0.8\textwidth]{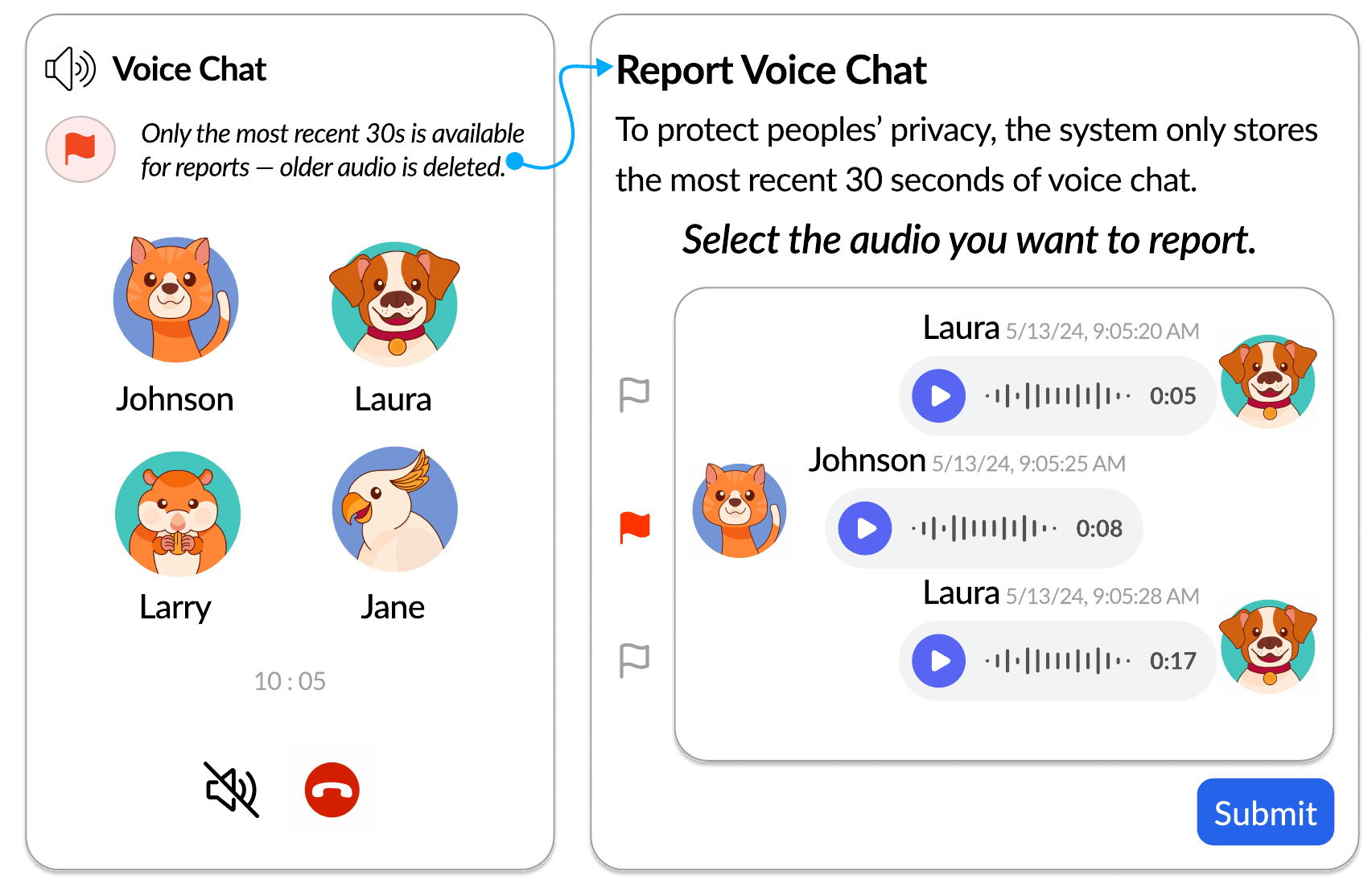}
    \caption{\textbf{Ephemeral Reporting for Voice Chats}. This figure illustrates how users can report content from ephemeral voice conversations. To respect privacy expectations, only recent audio segments are available for reporting, while older ones are automatically deleted. Each reporting system can configure how many seconds of audio remain reportable. Since user speech may overlap, the interface should indicate the speaker associated with each audio segment to support authentication.}
    \hfill
    \label{ephemeral_reporting}
\end{figure*}

\subsubsection{\textbf{Design Idea 1: Authenticated by Persisting Ephemeral Conversations}}
One potential solution is to record ephemeral conversations. 
For example, some Discord communities use bots to log every edited or deleted message into a dedicated moderation channel~\cite{kiene2019technological}.
However, if platforms or communities record ephemeral conversations without any constraints, it would fundamentally change their ephemeral nature and violate users’ privacy expectations~\cite{kotfila2014message}. 
Alternatively, the reporting user can choose to locally record the conversation, which may provide sufficient authentication if they follow recommended practices for client-side authentication.
M3 thus encouraged users to take screenshots: ``\textit{sometimes it is a lot easier to have screenshots, because they're permanent. So if any messages get deleted, you will still see like the full picture.}''
Still, users may realize the need to record only after they’ve received abusive messages, by which point critical context may already be lost.
In these cases, Discord moderators might simply ignore such evidence, as indicated by M4, ``\textit{If a message has the edited in its status, then that message generally will be ignored [and will not] be trusted ... [because] its content was not the original content in the message.}''

To address this issue, we propose an ephemeral reporting system that enables authentication of ephemeral messages while maintaining reasonable privacy guarantees (See Figure \ref{ephemeral_reporting}).
Specifically, the system would record a sliding window of recent conversation content, such as archived text messages, recorded voice chats, or video captures of VR interactions.
It could be hosted by platforms or community moderators, provided its implementation follows our proposed authentication practices.
When a user initiates a report, the system could authenticate any messages within that window. 
Since privacy expectations for ephemeral conversations vary across communities, the window size can be configured either by time (e.g., last 30 seconds) or by the number of recent messages (e.g., last 10 messages).

\subsubsection{\textbf{Design Idea 2: Authenticated by Testimonial Evidence}}
When ephemeral conversations are not persisted, users must rely on testimony from other conversation members to support authentication. 
Sometimes, users would initiate a report during the incident, hoping moderators can immediately join and witness unwanted behaviors~\cite{jiang2019moderation}. 
In other cases, they expect moderators to gather testimony from other conversation members or even the reported person to verify their descriptions of the conversation.
As described by M3, ``\textit{When someone was streaming incredibly NSFW things in the voice chat, we ended up messaging [these bystanders] and asking them what happened.}''
This is possible because Discord voice chats automatically log ``\textit{when someone joined and left voice chats, so [community moderators] can put together a timeline}'' (M1).
However, in private channels, such participant metadata is unavailable to community moderators~\cite{discord2025permissions}, who must instead rely on the reporting user to name potential witnesses. While this raises the risk that only sympathetic participants are selected, we did not encounter such abuse in our interviews.

\move{Move from Design Space about Collecting Testimony from Users}{A key design challenge is balancing the need for authentication with the reporting user’s privacy, particularly their desire not to be identified as the source of the report.
A straightforward design improvement would be to require moderators to obtain the reporting user’s consent and provide clear justifications before involving others in the investigation. 
On the other hand, when moderators feel that collecting additional testimony is crucial but the reporting user disagrees, moderators also have the right to terminate the report.
M5 expressed a similar idea, ``\textit{if you do not allow us to collect the information [from the reported person] we need to handle your case, your case is not going to be handled, because we're not going to make a under-informed decision.}}

\section{Discussion}\label{sec:discussion}
\revise{In this work, we use Discord communities as a case study to examine online reporting systems through an adversarial lens, identifying design opportunities that empower users while limiting potential abuse. In this section, we explore how these ideas might extend to platform-level reporting and integrate with cryptographic approaches amid growing distrust in platforms. We also highlight opportunities to adapt principles from offline legal systems.}

\subsection{Applying Adversarial Approaches to Platform-level Reporting Systems}
\revise{We focus on community-level reporting systems in this paper, in part because the internal workflows of platform moderation are rarely documented in rich details.
Still, users express similar frustrations with platform-level systems, many of which are around limited procedural fairness and weak privacy protections~\cite{kou2021flag, crawford2016flag, wang2023reporting}.
Our adversarial framework helps explain these concerns, for example, regarding how platforms manage documentary evidence.
Many platforms automate this process, assuming that the surrounding messages provide enough context for the reported incident.
For instance, League of Legends moderators are given access to the full in-game chat log~\cite{blackburn2014stfu}, while Google Chat and Facebook Messenger expose the 50 and 30 most recent messages, respectively~\cite{report2025google, report2025messenger}.
Although efficient, these methods risk both over-collection of unrelated or sensitive content and omission of key context from elsewhere.
Others, like Telegram, let users submit screenshots~\cite{report2025telegram}, shifting control to users but creating risks of manipulation or doubt over authenticity—even for users acting in good faith.
These extremes mirror trade-offs we analyzed in community systems, underscoring the need for more balanced, abuse-resistant approaches that still support user participation.}

\revise{Our proposed design space can also inform platform-level reporting systems by offering techniques that more intentionally balance privacy, accuracy, and accountability.
Some of our proposed design ideas, particularly those involving structured workflows for evidence authentication, can be automated and thus align well with the streamlined, scalable processes that platforms typically prioritize~\cite{seering2020reconsidering}. 
Examples include our design space for evidence authentication and ephemeral reporting.
However, other proposals introduce interaction models that are more difficult to apply at the platform level.
Designs like progressive disclosure or moderation by testimony involve multi-turn exchanges between moderators and involved users, which contrast with the one-shot, form-based reporting systems most platforms use today~\cite{crawford2016flag}. 
In addition, platform moderators are unfamiliar with the specific communities and may lack the context needed to conduct such deliberative investigations~\cite{seering2020reconsidering}. 
These challenges suggest the need for future work on how adversarial reporting mechanisms might scale under platform resource and workflow constraints.}

\subsection{Interplay between Cryptographic Tools and Adversarial Reporting Systems}
In our threat model, we assume that (1) users and moderators are UI-bounded in their ability to abuse the system, and (2) platforms are trusted to see user messages and operate without bias. 
However, these assumptions do not hold for all user populations.
Growing concerns over privacy and data misuse by platforms has led to the widespread adoption of E2EE messaging platforms such as WhatsApp, Signal, and iMessage~\cite{ermoshina2016end}.
Even Discord has begun rolling out E2EE for voice and video calls~\cite{discord2025e2eeav}.
Moreover, technically savvy users can modify client code to bypass interface constraints.
Modified versions of applications can be packaged and downloaded for easy use by others, in some cases eclipsing the popularity of the native app~\cite{munyendo2024you,doctorow2020eff}.

\revise{These trends complicate how some of our proposed designs can be implemented, but they also open up opportunities to implement them more securely through cryptographic techniques.
For example, many of our proposals (e.g., redaction tools and ephemeral reporting) implicitly assume that platforms can be the neutral and trusted intermediaries to host these reporting affordances.
If users who do not trust platforms wish to instead host these services on their local clients, moderators would require guarantees that their clients run the proper redaction algorithm or compute the correct response.
One approach is to require clients to generate cryptographic proofs of correct algorithmic execution~\cite{sun2024zkllm}.
Additionally, archiving a sliding window in ephemeral reporting might raise concerns over technical-savvy users that ``temporary'' conversations could be stored indefinitely.
To mitigate this, messages could be encrypted using time-limited keys that expire after a preset duration~\cite{specter2019keyforge, arun2022short, beck2023time}, ensuring that archived content becomes inaccessible after the intended retention window.
These examples illustrate how cryptographic tools can help strengthen the implementation of adversarial reporting systems under less trusted conditions.
We provide additional examples and highlight open research directions for applying cryptographic techniques to our design space in the Appendix~\ref{cryptographic_analysis}.}

\subsection{Applying Offline Legal Practices to Online Content Moderation}
As online spaces become an essential component of our society, the boundary between online content moderation and offline legal practices is increasingly blurred.
On one hand, this convergence opens up opportunities to draw from offline legal systems, which have long explored a wide range of procedural designs.
\revise{Beyond the adversarial–inquisitorial distinction, comparative legal literature offers additional dimensions for thinking about procedural design, such as the intended goals of a process~\cite{fuller1978forms, damaska1986faces}, and how participation and transparency are structured~\cite{mashaw1985due}.
These perspectives can help inform a more principled design of online moderation and reporting systems.}
Additionally, comparisons between common law and civil law traditions can inform how communities structure their governance policies~\cite{chen2023case}.
Many communities rely on detailed written policies~\cite{fiesler2018reddit, butler2008don}, resembling the legal codes of civil law systems.
But moderators often struggle to update these documents to reflect evolving norms.
Drawing from case law traditions, we suggest that incorporating prior moderation decisions as precedents could ease this burden, while also increasing transparency for community members.

On the other hand, researchers should also be mindful of some fundamental differences between online and offline settings.
For example, the United States enacted the Digital Millennium Copyright Act (DMCA) in 1998, allowing copyright holders to request the removal of infringing content from online platforms.
Victims of non-consensual intimate imagery often rely on this mechanism to take down their photos, but filing a DMCA report can unintentionally cause further harm~\cite{qiwei2024law}.
To comply with offline due process, platforms need to collect the victim’s personal contact information and forward the request to the accused, who may then identify the reporter and retaliate~\cite{take2024stoking, shouting2023pen}.
Similarly, users are often allowed to record their own online conversations without notifying other conversation participants.
Although online voice chats resemble offline conversations, some offline jurisdictions require the consent of all participants before a conversation can be recorded.
More research is needed to understand the factors driving these differences between online and offline practices, and to explore how offline legal frameworks might be adapted to better fit online contexts.

\section{Limitations and Future Work}
\revise{Our goal in this work is to map out a comprehensive design space for incorporating adversarial elements into community-level reporting systems, rather than to propose or evaluate a single, concrete system that embodies these practices end to end.
In practice, communities may adopt these designs selectively, or integrate individual components to address specific challenges within existing reporting workflows. 
Prior social computing research has similarly applied external theoretical frameworks—such as affirmative consent or restorative justice—to analyze online harm and reason about design possibilities without testing specific systems~\cite{im2021yes, xiao2023addressing}. Accordingly, while our contributions do not provide empirical evidence about the effectiveness of adversarial reporting designs, future research can work with online communities to implement and evaluate selected aspects of this design space across different governance structures, moderation norms, and privacy expectations.}
Additionally, because there is limited publicly available information about the workflows of platform-level moderators, our analysis primarily draws on community-level reporting systems.
Future research should investigate how the proposed design space can be adapted and integrated into platform-level moderation workflows.

\section{Conclusion}
\revise{Community-level reporting systems are central to resolving interpersonal conflicts and protecting users from harm in online spaces, particularly those with heightened privacy expectations~\cite{pfefferkorn2022content, kamara2022outside}.
However, the inherent power imbalances between moderators and reporting users contribute to recurring concerns about procedural justice and privacy risks~\cite{wang2023reporting, jiang2019moderation}. 
In this work, we apply adversarial legal frameworks to examine community-level reporting systems, using Discord as a research site. We show that these concerns can be traced in part to the inquisitorial structure of exiting reporting systems, in which moderators lead evidence collection and case development~\cite{damaska1986faces}. 
By contrast, adversarial models grant users greater control over evidence and can better support procedural justice and privacy, but also introduce new risks that require careful threat modeling~\cite{thibaut1973procedural, tyler2006people, damaska1972evidentiary}. Building on this analysis, we outline a design space for supporting evidence disclosure and authentication within the constraints of community moderation and current platform affordances. More broadly, this work illustrates how comparative legal perspectives can inform the procedural design of reporting systems, and how these ideas may extend to platform-level reporting and cryptographic approaches amid growing distrust in platforms.}


\begin{acks}
This work was funded in part by NSF grant CNS-2120651. We would like to thank the members of the Social Futures Lab at the University of Washington for their invaluable help in this project. We also would like to thank our anonymous reviewers for their insightful feedback. Finally, we would like to express our heartfelt thanks to all the participants who dedicated their time and effort to participating in our study.
\end{acks}

\bibliographystyle{ACM-Reference-Format}
\bibliography{privacyreporting}

\appendix
\section{Deeper Exploration of Cryptographic Tools for Adversarial Reporting}\label{cryptographic_analysis}
\move{From Discussion}{We consider an alternative threat model in our discussion where users do not trust platforms and might go beyond UI-bounded attacks to abuse the reporting system.
Fortunately, cryptographic researchers have already explored various techniques under this broader threat model. Our design space further points to how these techniques can be adapted and extended to build more adversarial reporting systems.}

\move{From Discussion}{The main challenge in this setting is in evidence authentication.
The platform cannot be relied on since it does not see the plaintext messages (only encrypted ciphertexts).
Users cannot be relied on either since non-cryptographic user-generated evidence, such as screen recording, can be easily forged, whereas UI-bound restrictions like forwarding messages can be manipulated using modified client applications.
To address this, cryptographers have developed special-purpose reporting protocols known as \emph{message franking}~\cite{messenger2017facebook,grubbs2017message, tyagi2019asymmetric} to work in tandem with E2EE messaging protocols to allow users to report messages to the moderator.
In short, message franking creates cryptographic digital signatures on messages with the unique property that they can only be verified by the conversation participants and the moderator, preserving privacy (via deniability) against other parties.}

\move{From Discussion}{Existing message franking protocols primarily focus on reporting at the ``message'' level; moderators can verify reports for single messages but do not have cryptographic guarantees about the larger conversation structure or the ordering of messages.
As discussed earlier, conversation structure and ordering is important in our proposals for providing moderators with enough context to make informed decisions and enable progressive disclosure.
Some prior work has considered selective reporting of parts within a message~\cite{chen2018people,leontiadis2023private}, but selective reporting of messages within a conversation while maintaining conversation structure is an open problem.}

\move{From Discussion}{Our proposals also consider the role of anonymity in the reporting process, both for the reporting user and for the moderators.
Anonymity is challenging in the cryptographic setting since it is seemingly at odds with authenticity: message reports must be verified as being authentic and sent by a real user.
Prior cryptographic message franking work has considered hiding the reporting user's identity~\cite{lai2023asymmetric} and revealing a reportee's identity only after a quorum of moderators agree on a decision~\cite{pattison2023committee}.
These protocols may be adapted and extended to consider the additional settings we propose, including partial moderator anonymity, user pseudonyms, or targeting a selected set of moderators.}

\move{From Discussion}{Lastly, a set of our design ideas also involve running client-side algorithms, for instance, to redact with custom attributes or answer more open-ended questions via an LLM.
To translate this class of ideas to the modified client threat model, moderators will need assurance that clients run the proper redaction algorithm or compute the correct response.
One direction would be to require clients to produce cryptographic proofs of correct execution of the algorithm~\cite{sun2024zkllm}.
Unfortunately, for complex algorithms, encoding them within a cryptographic proof will likely be prohibitively expensive.
In that case, one might rely on trusted execution environments~\cite{sabt2015trusted} in which a hardware component on a user's device attests to running the code client-side.}



\section{Literature Concepts}
Because our analysis draws heavily on legal literature, we summarize the key concepts from the literature that informed our work in Figure~\ref{framework}.

\begin{figure*}
    \centering
    \includegraphics[width=\textwidth]{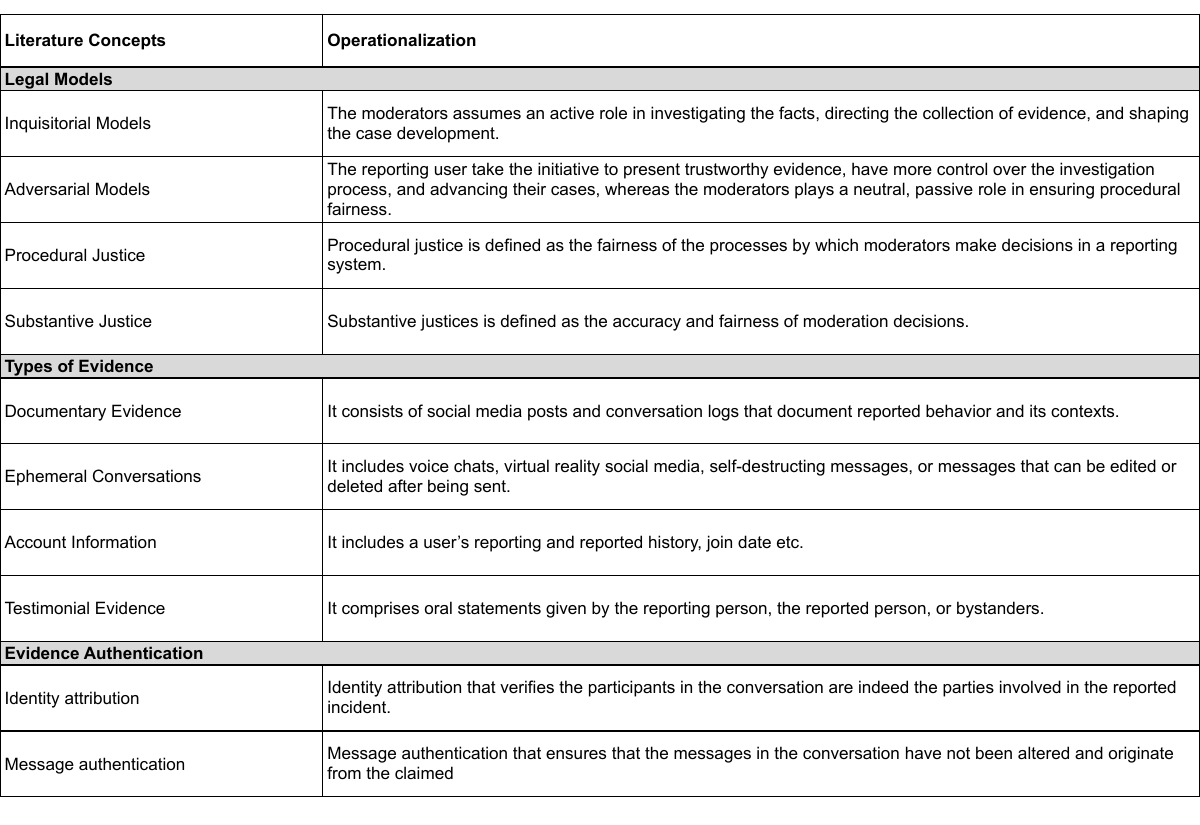}
    \caption{\textbf{Key literature-derived concepts.} 
    This figure highlights central constructs from legal scholarship (e.g., adversarial versus inquisitorial models, procedural justice) that guided our analysis.}
    \label{framework}
\end{figure*}

\section{Codes of Interpretive Analysis}\label{thematic-analysis}
\begin{figure*}
    \centering
    \includegraphics[width=\textwidth]{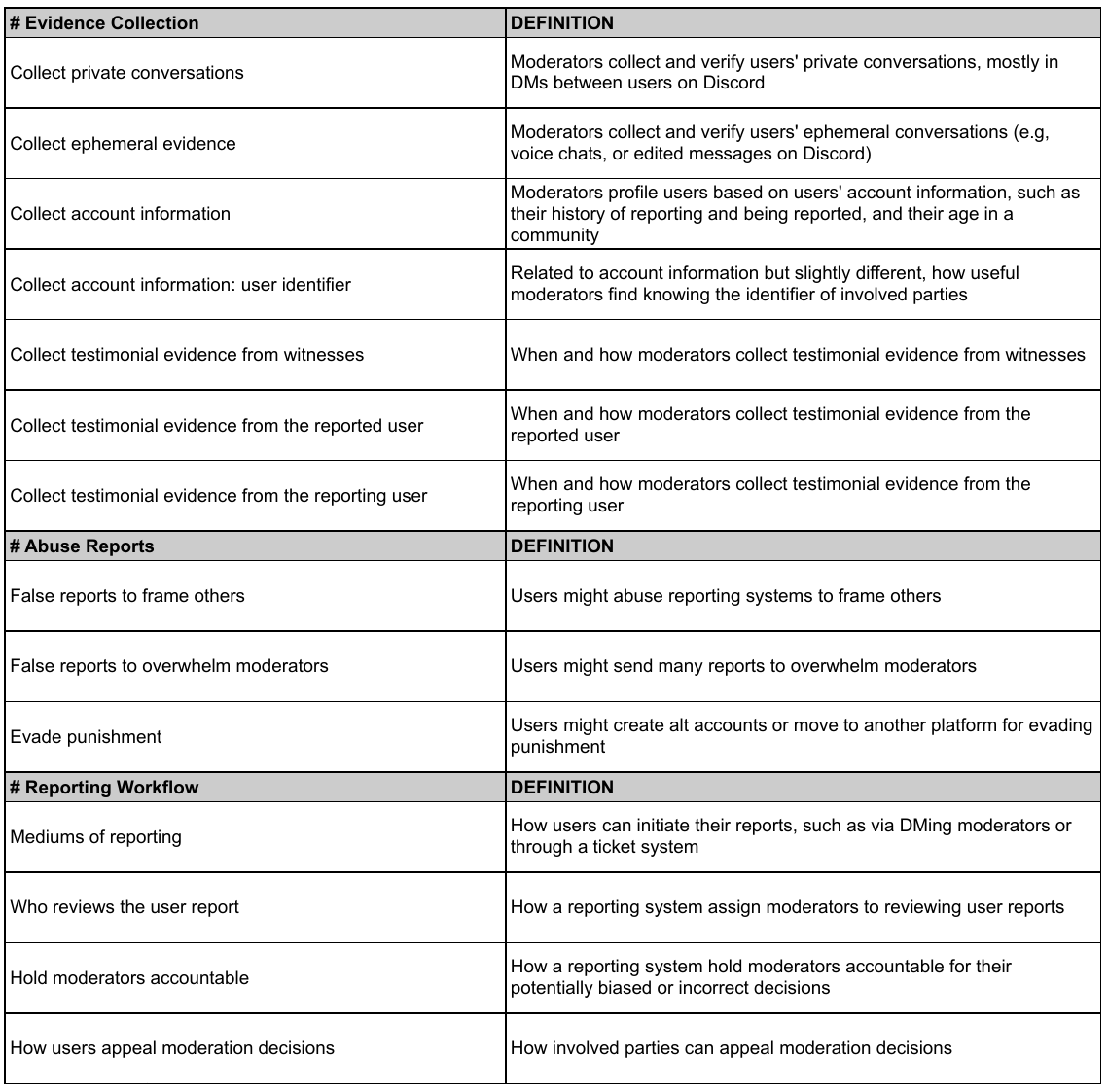}
    \caption{\textbf{Example codes from our interpretive analysis.} 
    These codes illustrate how interview data were coded and interpreted through concepts drawn from adversarial legal theory.}
    \label{thematic_analysis}
\end{figure*}

To support our interpretive analysis informed by adversarial legal theory, we coded interview transcripts using a combination of open coding and theory-guided interpretation. Drawing on concepts from adversarial legal literature, we applied this coding framework to examine reporting practices in Discord communities. Figure~\ref{thematic_analysis} presents a subset of example codes used in this analysis.

\end{document}